\documentclass[11pt,a4paper]{article}%
\pdfoutput=1

 \usepackage[normalem]{ulem}
\usepackage[T1]{fontenc}
\usepackage{amsmath}
\usepackage{amssymb}
\usepackage{amsthm}
\usepackage[ 
    ]{hyperref}
\usepackage{mathtools}
\usepackage{graphicx}
\usepackage{booktabs}
\usepackage{caption}
\usepackage{slashed}
\usepackage{listings}
\usepackage{slashed}
\usepackage{rotating}
\usepackage{subfig}
\usepackage{geometry}
\usepackage{multicol}
\usepackage[nottoc]{tocbibind}
\usepackage{multicol}
\usepackage{authblk}
\usepackage{cite}
\usepackage{appendix}
\usepackage{color}
\usepackage{epsfig}
\usepackage{hyperref}
\usepackage{enumerate}
\usepackage{amsfonts}
\usepackage{epstopdf}
\usepackage{latexsym}
\usepackage{arydshln}

\definecolor{myred}{rgb}{0.7, 0, 0}
\definecolor{myblue}{rgb}{0, 0, 0.7}
\definecolor{mygreen}{rgb}{0.04, 0.7, 0.5}

\hypersetup{colorlinks,citecolor=myred,linkcolor=myblue,urlcolor=myblue,linktocpage=true}

\newcommand{\be}{\begin{equation}}
\newcommand{\ee}{\end{equation}}
\newcommand{\bea}{\begin{eqnarray}}
\newcommand{\eea}{\end{eqnarray}}

\def\be{\begin{equation}}
\def\ee{\end{equation}}

\makeatletter

\@ifundefined{definecolor}
{\usepackage{color}}{}

\newcommand{\beq}{\begin{equation}}
\newcommand{\eeq}{\end{equation}}  
\newcommand{\ba}{\begin{eqnarray}}
\newcommand{\ea}{\end{eqnarray}}
\newcommand{\bef}{\begin{figure}}
	\newcommand{\eef}{\end{figure}}

\newcommand{\dn}{\Delta N_\text{eff} }

\definecolor{gbcolor}{rgb}{.2,.15,.82}
 
\definecolor{gbcolor1}{rgb}{.7,.15,.82}

\title{\Large \textbf{
The $H_0$ tension:  $\Delta G_N$ vs. $\Delta N_{\rm eff}$}}

\author{
{\large Guillermo Ballesteros$^{1, 2}$, Alessio Notari$^{3}$ and Fabrizio Rompineve$^{4}$}\\
\normalsize\itshape $^1$Instituto de F\'isica Te\'orica UAM/CSIC, Calle Nicol\'as Cabrera 13--15, Cantoblanco E-28049 Madrid, Spain\\
\normalsize\itshape $^2$Departamento de F\'isica Te\'orica, Universidad Aut\'onoma de Madrid (UAM)\\ 
\normalsize\itshape Campus de Cantoblanco, E-28049 Madrid, Spain\\
 \normalsize\itshape $^3$ Departament de F\'isica Qu\`antica i Astrofis\'ica \& Institut de Ci\`encies del Cosmos (ICCUB), Universitat de Barcelona, Mart\'i i Franqu\`es 1, 08028 Barcelona, Spain\\
\normalsize\itshape $^4$ Institute of Cosmology, Dept.~of Physics and Astronomy, \\
\normalsize\itshape Tufts University, Medford, MA 02155, USA 
}

\begin{document}
	
	\begin{flushright}
\footnotesize
{IFT-UAM/CSIC-20-54}
\end{flushright}

{\let\newpage\relax\maketitle}

	\begin{abstract}

\noindent We investigate whether the $4.4\sigma$ tension on $H_0$  between SH$_{0}$ES 2019 and Planck 2018 can be alleviated by a variation of Newton's constant $G_N$ between the early and the late Universe. This changes the expansion rate before recombination, similarly to the addition of $\Delta N_{\rm eff}$ extra relativistic degrees of freedom. We implement a varying $G_N$ in a scalar-tensor theory of gravity, with a non-minimal coupling of the form $(M^2+\beta \phi^2)R$.  If the scalar $\phi$ starts in the radiation era at an initial value $\phi_I \sim 0.5~M_p$ and with $\beta<0$, a dynamical transition occurs naturally around the epoch of matter-radiation equality and the field evolves towards zero at late times. As a consequence, the $H_0$ tension between SH$_{0}$ES (2019) and Planck 2018+BAO slightly decreases, as in $\dn$ models, to the 3.8$\sigma$ level. We then perform a fit to a combined Planck, BAO and supernovae (SH$_0$ES and Pantheon) dataset. When including local constraints on Post-Newtonian (PN) parameters, we find $H_0=69.08_{-0.71}^{+0.6}~\text{km/s/Mpc}$ and a marginal improvement of $\Delta\chi^2\simeq-3.2$ compared to $\Lambda$CDM, at the cost of 2 extra parameters. In order to take into account scenarios where local constraints 
could be evaded, we also perform a fit without PN constraints and find $H_0=69.65_{-0.78}^{+0.8}~\text{km/s/Mpc}$ and a more significant improvement $\Delta\chi^2=-5.4$ with 2 extra parameters. For comparison, we find that the $\Delta N_{\rm eff}$ model gives $H_0=70.08_{-0.95}^{+0.91}~\text{km/s/Mpc}$ and $\Delta\chi^2=-3.4$ at the cost of one extra parameter, which disfavors the  $\Lambda$CDM limit  just above 2$\sigma$, since $\Delta N_{\rm eff}=0.34_{-0.16}^{+0.15}$.
Overall, our varying $G_N$ model performs similarly to the $\dn$ model in respect to the $H_0$ tension, if a physical mechanism to remove PN constraints can be implemented.

\end{abstract}

\newpage
	

\section{Introduction}

The expansion rate of the Universe is currently at the center of an observational tension. On the one hand, the present Hubble parameter $H_{0}$ can be determined by measuring distances of astronomical objects according to the distance ladder method. The most precise recent measurement of this kind has been performed by the SH$_{0}$ES team  using Supernovae data and gives $H_{0}=74.03\pm 1.42~\text{km/s/Mpc}$~\cite{Riess:2019cxk}, with a calibration method based on Cepheids. On the other hand, $H_0$ is independently determined by Cosmic Microwave Background (CMB) data, once a given cosmological model is specified. The latest fit of the standard six-parameter $\Lambda$CDM model to CMB temperature, polarization and lensing power spectra measured by the Planck collaboration gives $H_{0}=67.27\pm 0.60$ km/s/Mpc~\cite{Aghanim:2018eyx}. Therefore, the two measurements disagree at 4.4$\sigma$. When combined with baryon acoustic oscillations (BAO) data, Planck finds $H_{0}=67.66\pm 0.42~\text{km/s/Mpc}$~\cite{Aghanim:2018eyx}, assuming the six-parameter $\Lambda$CDM. Other local measurements generally prefer values of $H_{0}$ which are higher than the CMB measurement. In particular, the measurement by the H0LiCOW team from quasar lensing is $H_0=73.3 \pm 1.8 \text{km/s/Mpc}$ \cite{Birrer:2018vtm}. An exception to this general split between local and early-time measurements is that of~\cite{Freedman:2019jwv}, which finds $H_0 = 69.8 \pm 0.8 ({\rm stat}) \pm 1.7 ({\rm sys}) $ km/sec/Mpc, midway in the range defined by Planck and SH$_0$ES. For recent reviews on the current observational status of the tension see~\cite{Verde:2019ivm} and \cite{Riess:2020sih}.

At the time of writing, no satisfactory explanation of the discrepancy between the various measurements based on systematic errors has emerged.  While we wait for the final observational verdict, it is interesting to ask whether a cosmological model different from $\Lambda$CDM can resolve the tension between empirical and cosmological model-dependent determinations of $H_{0}$.
Different efforts in this direction can be broadly classified as late or early time attempts
(see~\cite{Knox:2019rjx} for a comprehensive review): the former modify the cosmological history only much after recombination, whereas the latter feature changes before or around recombination. 

The Planck and SH$_0$ES measurements of $H_0$ can be made to agree better by  changing the expansion history after recombination, with a different time evolution of the angular diameter distance with respect to standard $\Lambda$CDM. However, BAO data is in conflict with this kind of solution, which makes early time models the most effective strategy to solve the $H_0$ tension, see e.g.\ \cite{Aylor:2018drw}. Arguably, the simplest such attempt consists in adding one additional parameter to the base six-parameter $\Lambda$CDM,  allowing for extra relativistic species ($\Delta N_{\rm eff}$) beyond the Standard Model neutrinos, e.g. axions~\cite{DEramo:2018vss}, non-thermal light dark matter \cite{Alcaniz:2019kah} and gravitational waves (see e.g.~\cite{Graef:2018fzu}), among other candidates. This modification, shortly the `$\Delta N_{\text{eff}}$ model', alleviates the tension~\cite{Aghanim:2018eyx,DEramo:2018vss}, but does not solve it completely, mainly because the required value of $\Delta N_{\text{eff}}$ also affects the photon diffusion scale and thus spoils the fit to the CMB damping tail. We will nonetheless consider such a model using recent 2019 SH$_{0}$ES and Pantheon data,
showing a rather significant shift of the fit with respect to that reported from the old 2018 SH$_{0}$ES data \cite{Aghanim:2018eyx}. Moreover, the $\Delta N_{\rm eff}$ model constitutes a useful benchmark to compare how well other theoretical proposals perform.

A better fit to Planck, BAO and SH$_{0}$ES data than that of the six-parameter $\Lambda$CDM model was shown to be provided by the addition of an extra early dark energy (EDE) component, which contributes to $\sim5\%$ of the total energy density of the Universe just before the epoch of matter-radiation equality~\cite{Poulin:2018cxd,Smith:2019ihp,Lin:2019qug}, and then dilutes faster than radiation, in such a way as to minimize the effects on the photon diffusion scale. Field theory realizations of this idea~\cite{Poulin:2018cxd,Smith:2019ihp} employ a light scalar field, which is initially frozen in its potential due to Hubble friction. Once the Hubble rate drops to values comparable to the curvature of the potential, the field starts rolling and may or may not oscillate, depending on the properties of the potential and the initial field value. Simple power-law potentials fit the data only marginally better than the $\Delta N_{\text{eff}}$ model, with $\phi^{4}$ being the preferred potential~\cite{Agrawal:2019lmo} (in this case the first few oscillations provide a short epoch where $w\gtrsim 1/3$). To date, the best fit to the CMB, BAO and SH$_{0}$ES data sets (in this kind of models) is provided by a scalar field which initially sits in the concave region of a $\cos(\phi)^{n}$ potential, with $n>1$ ($n=3$ being the preferred power)~\cite{Poulin:2018cxd, Smith:2019ihp}. See however~ \cite{Hill:2020osr} for a recent critical take on EDE models when Large Scale Structure (LSS) data are taken into account.

Despite their relative success in alleviating the Hubble tension, the above scenarios may still be considered unattractive compared to the simpler $\Delta N_{\text{eff}}$ extension of $\Lambda$CDM, for the following reasons: Firstly, none of them explains why the transition in the EDE component occurs around matter-radiation equality. Rather, the curvature of the potential is fixed ad-hoc to be of the order of the Hubble rate at the relevant epoch. Therefore, these scenarios suffer from a coincidence problem. 
Secondly, the EDE scenario of~\cite{Poulin:2018cxd} requires a potential which, although periodic, does not match the standard potential for axion-like fields, and whose field theory origin 
may thus be considered uncertain. Furthermore, the latter model introduces four extra parameters with respect to the base six-parameter $\Lambda$CDM. 

It is this unsatisfactory situation which we take as motivation for this work. We aim at finding a model alternative to $\dn$, with the smallest number of extra parameters and which does not suffer from a coincidence problem. Our approach stems from the realization that the background effect of dark relativistic species in the early Universe, e.g.\ at the epoch of Big Bang Nucleosynthesis (BBN), can be mimicked by a varying Planck mass in a Universe with the standard matter and radiation content, see e.g.\ \cite{Iocco:2008va}. That this is indeed the case can be easily understood by looking at Friedmann's equation: on the one hand in the $\dn$ model the additional energy density in dark radiation (and in the other species, including dark matter, to keep the fit to CMB data) necessarily corresponds to an increase in the Hubble parameter; on the other hand the same increase can be obtained by keeping the standard radiation and matter content but taking the Planck mass to be smaller in the early Universe than it is today. The extent of the analogy between the latter and the former models as we approach the epoch of matter-radiation equality is less clear, in particular with respect to the Hubble tension. Indeed, as we will show, the shift in the total energy density caused by a varying $G_N$ can scale differently from radiation at this epoch. Furthermore, the behavior of cosmological perturbations is in general different in the two models. In this work, we would therefore like to assess whether a varying Newton's constant can alleviate the Hubble tension, as an alternative scenario to the $\Delta N_{\text{eff}}$ model.

In order to do so, we will consider one of the simplest and popular implementations of this idea in field theory. This is provided by a scalar field non-minimally and quadratically coupled to gravity, similar to the old Brans-Dicke proposal~\cite{Brans:1961sx}. The model which we will focus on adds only two extra parameters to the $\Lambda$CDM model, which correspond to the minimum number of parameters needed to describe a time varying Newton's constant: its initial value in the early Universe and the rate of its variation. Very interestingly, this field theory scenario presents some of the ingredients that characterize the aforementioned oscillating scalar field models, without suffering from the same coincidence problem. Indeed, an essential feature of the epoch of matter-radiation equality is a change in the evolution of the Hubble parameter. As a consequence of Einstein's equations, the gravitational background field also changes at matter-radiation equality: in particular, the Ricci scalar goes from being approximately vanishing during radiation domination to a non-zero value during matter domination. Therefore, by coupling a scalar field to the Ricci scalar, a dynamical transition at the epoch of matter-radiation equality arises naturally. 
While this is equivalent to a model with canonical Einstein action but with the scalar field coupled directly to the matter Lagrangian, it is easier to understand its most interesting aspects in the so-called Jordan frame, where the scalar field has a time dependent mass which is proportional to the Ricci scalar.

If we set the scalar potential to zero, the field is essentially frozen during radiation domination, and becomes dynamical only close to the onset of matter domination. Depending on initial conditions and on the value of the (dimensionless) non-minimal coupling, the field can then roll or oscillate around matter-radiation equality.
Its energy density then redshifts faster than radiation and makes the scenario promising from the point of view of the Hubble tension.
By performing a fit to cosmological data, we confirm that a larger Hubble constant can indeed be accommodated in our setup. The likeliness of this scenario in comparison to that of the $\Delta N_{\text{eff}}$ and $\Lambda$CDM models depends on the behavior of the non-minimally coupled scalar field in the Solar System, where modifications to General Relativity are strongly constrained. Nevertheless, scenarios exist where such constraints 
could be evaded~(e.g.~by means of screening mechanisms such as~\cite{Vainshtein:1972sx, Khoury:2003aq}). We thus perform two analyses, with and without including local constraints on Post-Newtonian parameters.

Before moving to the main body of the paper, let us mention previous work in similar directions. Implications of early modified gravity scenarios for the Hubble tension have been investigated in~\cite{Lin:2018nxe}, albeit by means of parametrized phenomenological deviations from General Relativity, rather than in a field theory model. CMB bounds on a field theory scenario which is somewhat similar to ours have been presented in~\cite{Rossi:2019lgt}, where the relation to the Hubble tension was also investigated, albeit a fit with SH$_{0}$ES data was performed only for a specific choice of the non-minimal coupling, i.e.\ the conformal case. See also \cite{Umilta:2015cta, Ballardini:2016cvy} for earlier, related works. Similar ideas to ours have been also recently discussed in \cite{Sola:2019jek}, in the context of the standard Brans-Dicke theory (i.e.\ without a constant mass parameter in the Einstein-Hilbert action) and neglecting local constraints on deviations from General Relativity, as well as in~\cite{Zumalacarregui:2020cjh}, in frameworks with more extra parameters than the model which we focus on here. We comment further on the relation of our work to~\cite{Rossi:2019lgt} and~\cite{Zumalacarregui:2020cjh}  below. Finally, non-minimal couplings to alleviate the Hubble tension have also been considered in~\cite{Sakstein:2019fmf}, which however makes use only of threshold effects on the scalar field, due to neutrinos becoming non-relativistic. Such threshold effects are instead subdominant in our scenario. More broadly, the implications of other varying fundamental constants for the Hubble tension has also been recently investigated in~\cite{Hart:2019dxi}.

Our paper is organized as follows. In section~\ref{Model} we present the non-minimally coupled scalar field model; in section~\ref{Results} we present the results of the fits to cosmological data and in section~\ref{Conclusions} we draw our conclusions. We use natural units throughout the paper.

\section{Model} \label{Model}

In order to investigate the implications of a varying $G_N$ scenario for the Hubble tension, we will consider a simple scalar-tensor model which introduces only two new parameters with respect to $\Lambda$CDM. It is easy to see that this is the minimum number of extra parameters which is needed to capture variations of $G_{N}$: one parameter corresponds to the difference between the values of the Newton's constant in the early Universe and today; a second parameter is needed to describe the rate of variation of $G_N$. This is in contrast with the $\Delta N_{\text{eff}}$ proposal to address the Hubble tension, which introduces only one extra parameter to $\Lambda$CDM.\footnote{However, again at least two parameters are needed to describe new species which behave as radiation at early times, with a change in the equation of state parameter at late times.} More complicated models, with more than two new parameters, can also be considered (see e.g.~\cite{Zumalacarregui:2020cjh} for recent work in the context of the Hubble tension) and we will briefly comment on this possibility later on.
	
The simplest scenario is a modification of Einstein's gravity, obtained by coupling a scalar field $\phi$ that sets the value of $G_N$ to the Ricci scalar $R$.  The action of this non-minimally coupled scalar, in the so-called Jordan frame, is
\begin{eqnarray}
\label{eq:nmaction}
S&=& \frac{1}{2}\int d^4x \sqrt{-g}\left[M^2 f(\phi)  R+\partial_\mu \phi \partial^\mu \phi +L_{\rm tot} \right] \, ,
\end{eqnarray}
where
\begin{eqnarray}
 f(\phi) \equiv 1+ \beta \frac{\phi^2}{M^2} \, ,
\end{eqnarray}
$\beta$ is a dimensionless coupling constant assumed to be negative, $M$ is a constant mass scale and $L_{\rm tot}$ represents the remaining contents of the Universe, including radiation, dark matter, baryons, neutrinos and a cosmological constant $\Lambda$. We assume a negligible mass in the Jordan frame for $\phi$. The same model can be presented in the so-called Einstein frame by means of a Weyl transformation of the metric tensor $g^{E}_{\mu\nu}=f(\phi)g_{\mu\nu}$. Then the Ricci term in the action becomes canonical, but the field $\phi$ couples directly to the matter Lagrangian. Physically measurable quantities are of course the same in the two frames. 

The background equation of motion of the scalar in a flat FLRW metric in the Jordan frame is
\begin{eqnarray}
\ddot{\phi}+3 H\dot{\phi}-\beta\, R\, \phi=0 \, , \label{eom}
\end{eqnarray}
where a dot denotes a derivative with respect to cosmic time $t$, $H\equiv {\dot{a}}/{a}$ is the Hubble parameter and $a(t)$ is the FLRW scale factor. The Ricci scalar can be expressed as a function of $H$ as  
\begin{equation}
R=6 \dot{H}+12 H^2.
\end{equation} 
In the deep radiation era, $H(t)\simeq 1/(2t)$, thus $R\approx 0$ and the field is frozen at some initial value $\phi_I$, which can be expected to be of the same order of $M$.\footnote{We neglect threshold effects due to particles that become non-relativistic~\cite{Damour:1994zq} and the conformal anomaly due to the running of coupling constants~\cite{Cembranos:2009ds}. The first effect has been used in~\cite{Sakstein:2019fmf} to address the $H_0$ tension. Both effects are subdominant in our scenario.} By writing Friedmann's equation as $3 H^2 M^2=\rho_\phi+\rho_{\rm tot}$, one finds
\begin{equation}
\label{eq:endensity}
\rho_{\phi}=\rho_{1}+\rho_{2}+\rho_{3}=\frac{1}{2}\dot{\phi}^2-6\beta H \phi\dot{\phi}-3\beta H^2\phi^2,
\end{equation}
which may be interpreted as an energy density of the homogeneous scalar field in this frame. Here $\dot\phi^2/2$ is the kinetic energy of the scalar field, and the last term is a shift in the usual critical energy density $3H^2 M^2$, since in our case the effective Planck mass is field dependent: $M^2_{*}(\phi)\equiv M^2+ \beta \phi^2$. Since $\beta<0$, the Planck mass is smaller (and thus gravity is stronger) in the early Universe than it is today. In order to have always a positive $M_*^2$ we will impose $\phi_I^2< M^2/\beta$.

As the Universe approaches the matter dominated era, $R>0$, the scalar field acquires an effective mass squared of order $\beta H^2$ and starts rolling towards zero reaching a final value $\phi_0$, typically much smaller than $\phi_I$. Therefore, the dynamics naturally features the ``release'' of an initially frozen scalar around the epoch of matter-radiation equality. This is in stark contrast to usual EDE models, where the time of transition from $w=-1$ to $w \geq 1/3$ has to be set in an ad-hoc manner to address the Hubble tension.

We will investigate the transition from the radiation dominated era to the matter dominated epoch numerically. However, one of the most interesting features of our setup can be understood analytically, by solving \eqref{eom} in the matter dominated era. Setting $a\propto t^{2/3}$, one straightforwardly finds
\begin{eqnarray}
\phi\propto t^{\pm\frac{1}{2} \sqrt{1+\frac{16}{3}\beta}-\frac{1}{2}}\propto a^{\pm \frac{3}{4}\sqrt{1+\frac{16}{3}\beta}-\frac{3}{4}} \, .
\end{eqnarray}
Therefore, for $\beta<-3/16$ the field undergoes damped oscillations, while for $-3/16\leq \beta< 0$ the evolution towards zero is monotonic. Since $H\sim a^{-3/2}$, it is straightforward to check that all terms in \eqref{eq:endensity} scale as $a^{-4.5}$ when $\beta\leq-3/16$, averaging over oscillations when they are present. Therefore, the energy density of the background scalar field is diluted faster than radiation once matter dominates.
The scaling of \eqref{eq:endensity} at early times, during radiation domination, is  found by setting $\phi\sim \text{const.}$, so that only the last term in \eqref{eq:endensity} contributes and scales as $\sim H^2 \sim a^{-4}$, i.e. as radiation.

As noted in~\cite{Poulin:2018cxd} a new species that dilutes faster than radiation after equality might fit the CMB and supernovae data better than the $\Lambda$CDM model with the addition of dark relativistic species. In our setup this happens as long as $\beta\leq -3/16$.

\begin{figure}[t]
	\includegraphics[width=\textwidth]{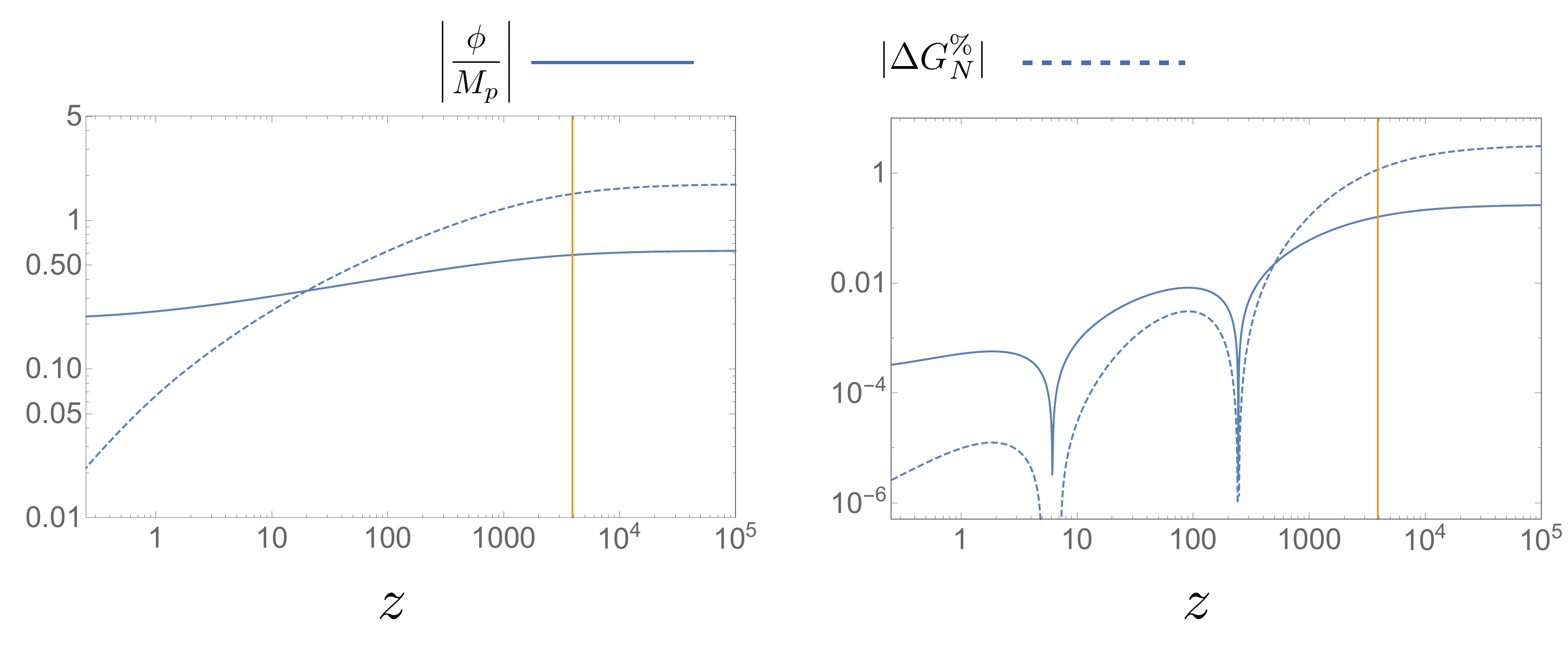}
	\caption{\small Evolution of the scalar $\phi$ in units of $M_p$ (solid blue) and the quantity $\Delta G_\%\equiv 100 |1-G_N/G_N(t_0)|$ (dashed blue) as a function of the redshift $z$. On the left: $\beta=-0.05$ and $\phi_{I}\simeq 0.7 M_p$. On the right: $\beta=-0.45$ and $\phi_{I}\simeq 0.3~M_p$. The orange vertical line denotes the redshift of matter-radiation equality. These plots are obtained using the best-fit values reported for our model in Table~\ref{tab:cosmoparametersl}: the left plot corresponds to the column `w/o PN', whereas the right plot corresponds to `w/ PN'.}
	\label{fig:both}
\end{figure}
However, the scalar-tensor model which we analyze in this paper is subject to two additional constraints. The first one applies to any model which predicts a variation of Newton's constant from the early Universe to today, since a too large deviation would spoil the agreement with Big Bang Nucleosynthesis (BBN) data. Secondly, at late times the simple model~\eqref{eq:nmaction} effectively introduces a fifth force, whose strength is severely constrained by local gravitational bounds on Post-Newtonian (PN) parameters. We discuss both constraints and their implications on the parameter space of the model \eqref{eq:nmaction} in detail in Sec.~\ref{Results}. However, let us point out here that local modifications to General Relativity can be suppressed in more complicated models, possibly with more than two parameters, where for instance the scalar field Lagrangian contains higher derivative terms which effectively screen the fifth force (see e.g.~\cite{Zumalacarregui:2020cjh} for a recent discussion).

\begin{figure}[t]
\centering
	\includegraphics[width=\textwidth]{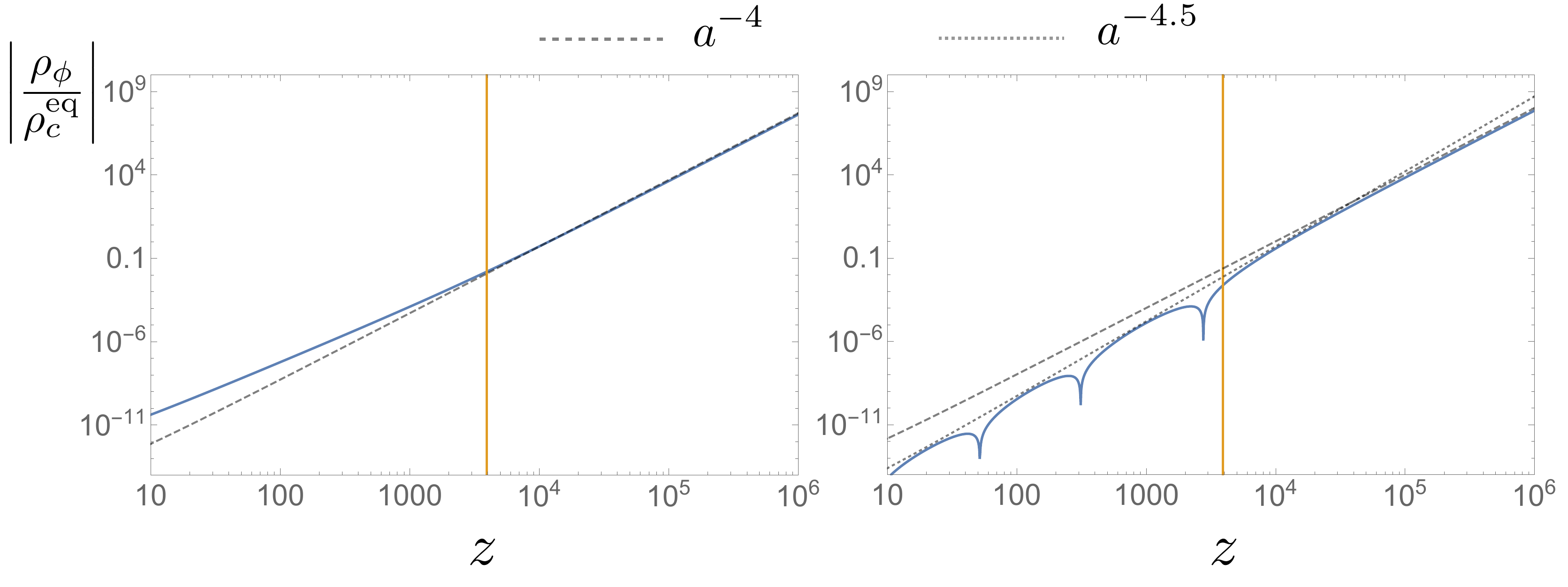}
	\caption{\small Contribution to the background energy density due to the scalar field, according to \eqref{eq:endensity}, as a function of $z$, normalized to the total energy density at matter-radiation equality. These plots are obtained using the best-fit values reported for our model in Table~\ref{tab:cosmoparametersl}: the left plot corresponds to the column `w/o PN', whereas the right plot corresponds to `w/ PN'. The orange vertical line denotes the redshift of matter-radiation equality. The dashed and dotted lines show different scalings of $\rho_\phi$ with $a$.}
	\label{fig:en}
\end{figure}

A detailed analysis of the dynamics and impact of the non-minimally coupled scalar requires the implementation of the model \eqref{eq:nmaction} in a Boltzmann code. We have done so by modifying {\tt{hi-class}}, a public code for scalar-tensor theories~\cite{Zumalacarregui:2016pph,Bellini:2019syt}, based on the Boltzmann code {\tt{CLASS}}~\cite{Blas_2011}. In the notation of {\tt{hi-class}}, the model \eqref{eq:nmaction} corresponds to  setting the functions $G_2= -\Lambda + X \equiv -(\Lambda+\partial_\mu \phi \partial^\mu \phi)/2$ and $G_4=M_*^2/2$, with and $G_3=G_5=0$. Furthermore, we measure mass scales and field values in units of the local effective Planck mass today, that is we set $M_p=(8\pi G_N^0)^{-1/2}=1$. See \eqref{eq:cavendish} for the definition of the effective Newton constant $G_N^0$, whereas we define $G_N(t)\equiv (8\pi M_\star(\phi))^{-2}$. The need to modify {\tt{hi-class}} arises because the original code does not support oscillating scalar fields, i.e.~can only deal with the monotonically rolling case and thus does not allow for a full exploration of the relevant parameter space.  In particular, our modifications to the code concern the evolution of the scalar field perturbations $\delta\phi(t,{\bf x})$, which are coupled to matter and metric perturbations and can thus crucially affect the CMB and the matter power spectra. The relevance of perturbations from the point of view of alleviating the Hubble tension, in particular for oscillating scalar fields, has been recently stressed in~\cite{Poulin:2018cxd}.  While the original {\tt{hi-class}} employs the variable $V_x\equiv -a \, \delta\phi(t,{\bf x})/\dot{\phi}$ for the equations of the perturbations, our modified code works directly with $\delta\phi(t,{\bf x})$ and it is thus well-behaved at the turning points of the background field. 

\begin{figure}[t]
\centering
	\includegraphics[width=\textwidth]{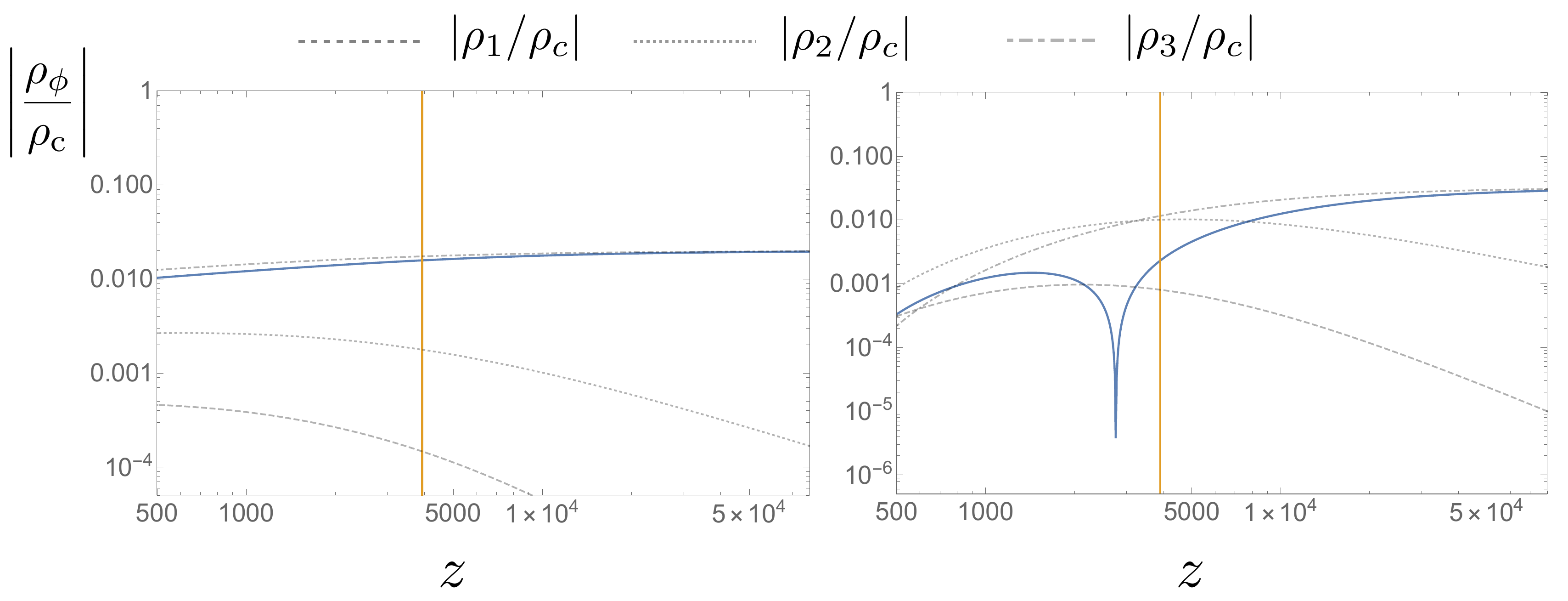}
	\caption{\small Ratio of the energy density due to the scalar field to the total energy density, as a function of $z$. These plots are obtained using the best-fit values reported for our model in Table~\ref{tab:cosmoparametersl}: the left plot corresponds to the column `w/o PN', whereas the right plot corresponds to `w/ PN'. The orange vertical line denotes the redshift of matter-radiation equality. The dashed, dotted and dot-dashed lines correspond to the different contributions to $\rho_\phi$ in \eqref{eq:endensity}.}
	\label{fig:enzoom}
\end{figure}

We plot in Fig.~\ref{fig:both} the evolution of $\phi$ (absolute value) as well as of $|\Delta G_N|\equiv |1-G_N/G_N(t_0)|$ (being $G_N(t_0)$ the present cosmological value of $G_N$) as a function of redshift for two representative examples with $\beta\simeq -0.05$ and $\beta\simeq -0.45$. These values of parameters actually correspond to the best-fit results for our scenario, which we present in Table~\ref{tab:cosmoparametersl} (on the left/right, the best-fit without/with constraints on the PN parameter). Other cosmological parameters have also been fixed according to Table~\ref{tab:cosmoparametersl}. For the same choices we also plot in Fig.~\ref{fig:en} the total energy density of the scalar field, according to \eqref{eq:endensity}, as a function of redshift, normalized to the total energy at matter-radiation equality. Deep in the radiation era the field behaves as a fluid which tracks the radiation background, whereas after matter-radiation equality its energy density can dilute either slower or faster than radiation, with equation of state parameter $w\approx 1/2$ if $\lvert \beta \rvert$ is large enough. In Fig.~\ref{fig:enzoom}, we show the ratio of $\rho_{\phi}$ to the total energy density as a function of redshift. For completeness, we also show the different contributions to $\rho_{\phi}$: it is clear that at early times the term $3\beta H^2\phi^2$ dominates over the other terms in \eqref{eq:endensity}, whereas around and after matter-radiation equality the other terms can be equally relevant. These figures are produced using our modified version of {\tt hi-class}.

Changing the parameters $\beta$ and $\phi_I$ leads to variations in the redshift at which the field is released as well as in its contribution to the total energy density. In order to potentially alleviate the Hubble tension, the release should occur slightly before matter-radiation equality, while the scalar field energy density should be $O(5-10)\%$ of the background density at matter-radiation equality. The task of determining exactly which values of parameters lead to the best fit to cosmological data can be performed with a Monte Carlo analysis, whose results we report in the next section.

\section{Datasets and Results} \label{Results}

The data sets that we consider include the
 SH$_{0}$ES 2019 measurement of the present day Hubble rate $H_0 = 74.03\pm 1.42$ km/s/Mpc~\cite{Riess:2019cxk}, the Pantheon supernovae dataset (Pantheon)~\cite{Scolnic:2017caz}, Planck 2018
high-$\ell$ and low-$\ell$ TT, TE, EE and lensing data~\cite{Aghanim:2019ame}.  We also include BAO measurements from 6dFGS at $z = 0.106$~\cite{Beutler:2011hx}, from the MGS galaxy sample of SDSS at $z = 0.15$~\cite{Ross:2014qpa}, and
  from the CMASS and LOWZ galaxy samples of BOSS DR12 at $z = 0.38$, $0.51$, and $0.61$~\cite{Alam:2016hwk}.
We perform the analysis using the public code {\tt Monte Python}~\cite{Audren:2012wb}. 
We model neutrinos, using the standard treatment of the Planck collaboration, as two massless 
and one massive species with $m_\nu = 0.06$ eV [49]. We study our model by trading the parameters $\beta$ and $\phi_I$ for $\beta \phi_I^2$ and $\phi_I$, on which we apply flat priors.

Further relevant constraints on scalar-tensor models come from BBN and from Solar System tests of General Relativity. The latter in particular provide a stringent bound on the so-called Post-Newtonian (PN) parameter $\gamma_{\text{PN}}$ (see e.g.~\cite{EspositoFarese:2004cc}), which in our setup is predicted to be 
\begin{eqnarray}
\gamma_{PN}-1\equiv-\frac{f'(\phi_\text{loc})^2}{f(\phi_\text{loc})+2 f'(\phi_\text{loc})^2} \approx -4 \beta^2 \frac{\phi^2_\text{loc}}{M^2} \, ,\label{PN}
\end{eqnarray}
where $\phi_{\text{loc}}$ is the value of $\phi$ in the Solar System today and we have considered only the first non-trivial order in $\phi_{\text{loc}}/M$ in the last expression. We will consider two possibilities in this work: first, we identify $\phi_{\text{loc}}$ with the average cosmological value of the scalar field, $\phi_0$, which is the output of the Boltzmann code. In this case, we thus apply existing constraints on $\gamma_{\text{PN}}$ in our MCMC runs. The most recent bound comes from the Cassini mission and is given by $\gamma_{PN}-1 =(2.1\pm 2.3)\times 10^{-5}$~\cite{Bertotti:2003rm, Fienga:2014bvy}, which we have included as a Gaussian constraint in our likelihood.
As an alternative possibility to include scenarios where the PN constraints could be evaded, for instance due to screening effects, we perform MCMC runs without 
any constraints on $\gamma_{PN}$. We refer to these two possibilities as `w/ PN' and `w/o PN' respectively.

\begin{figure}[t]
\vspace{-2cm}
	\includegraphics[width=\textwidth]{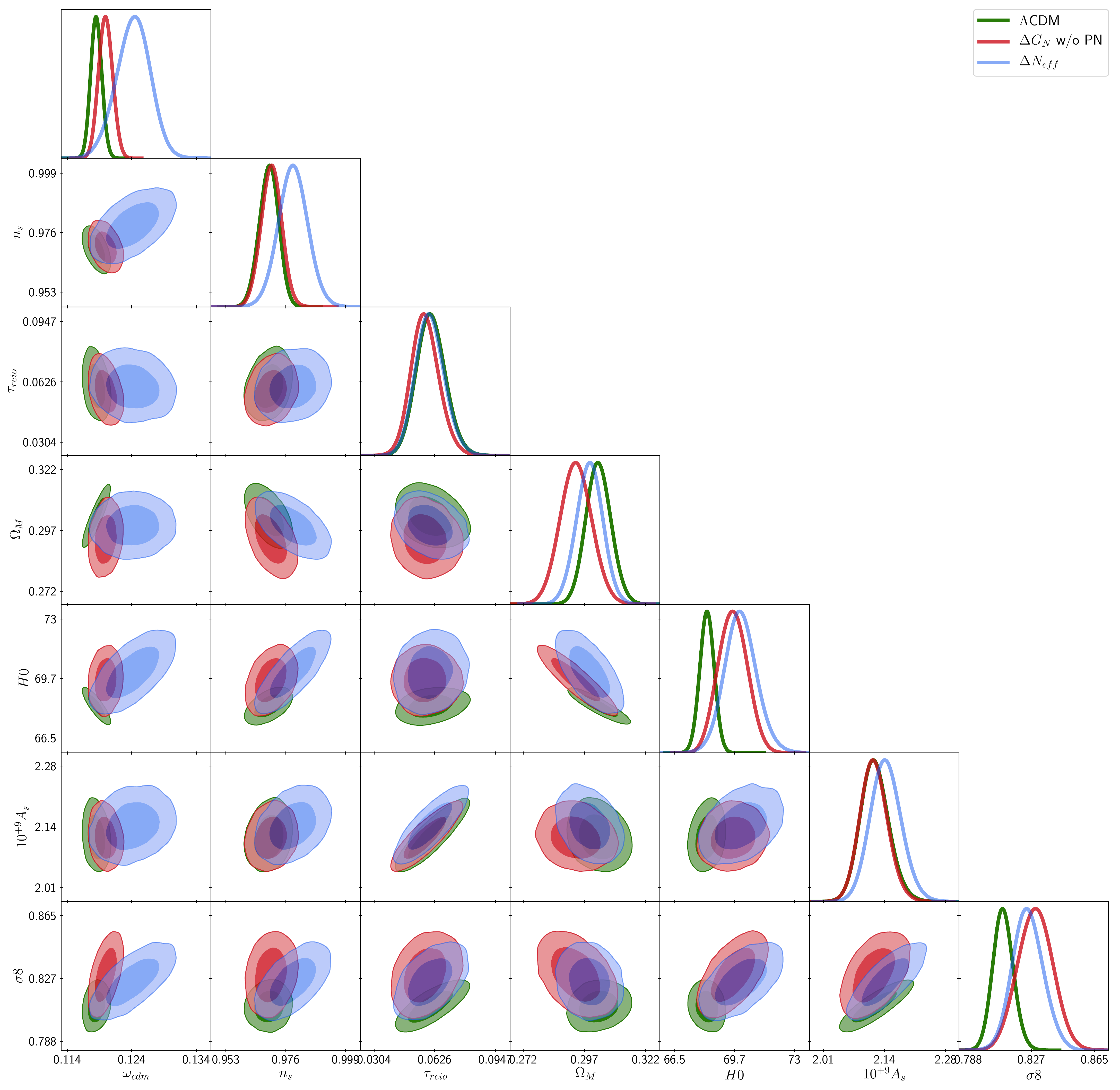}
	\caption{\small Constraints on parameters for our $\Delta G_N$ model vs.~the $\Delta N_{\rm eff}$ model and the base $\Lambda$CDM model, using
Planck 2018 high$-\ell$ TT,TE,EE+low$-\ell$ EE+ low$-\ell$ TT+lensing, BAO, Pantheon and SH$_{0}$ES 2019 data, without PN constraints. 
Parameters are our sampled MCMC parameters with flat priors. In particular a prior range has been set for the extra parameters: $-0.95<\beta \phi_I^2<0$
and $0<\phi_I<0.95$. Here $H_0$ is in km/s/Mpc. 
Contours contain $68 \%$ and $95 \% $ of the probability.  \label{triangleplot2}}
\end{figure}

The variation of the Newton constant from BBN until today is also constrained. In our setup, at the BBN epoch we have $G^{\text{BBN}}_N=G_N^I=1/(8 \pi M_*(\phi_I)^2)$, whereas the effective gravitational constant measured today in Cavendish-like experiments is given by~\cite{Boisseau:2000pr}:\footnote{As mentioned above, we set $(8\pi G_N^0)^{-1/2}=1$ in {\tt hi-class}. This is done under the assumption $\phi_\text{loc}=\phi_0$. However, we checked that setting instead $(8\pi G_N(t_0))^{-1/2}=1$ does not significantly alter our results.}
\begin{align}
\label{eq:cavendish}
 G_N^0 =\frac{1}{8\pi M^2 f(\phi_\text{loc}) }\left(\frac{2f(\phi_\text{loc})+4M^{2}f'(\phi_\text{loc})^2}{2f(\phi_\text{loc})+3M^{2}f'(\phi_\text{loc})^2}\right)
\approx \frac{1}{8\pi M^{2}}\left(1-\beta\frac{\phi_\text{loc}^{2}}{M^{2}}\right) \, ,
\end{align}
at the lowest order in $\phi_\text{loc}/M$. Neglecting $(\phi_\text{loc}/M)^2$ corrections our model predicts simply 
\begin{eqnarray}
\frac{G^{\text{BBN}}_N}{G^0_N} \approx 1-\beta \phi_I^2 \label{GBBN} \, .
\end{eqnarray}
To the best of our knowledge, the most conservative constraint from BBN is $G^{\text{BBN}}_N/G^0_N=1.01^{+0.20}_{-0.16}$~\cite{Copi:2003xd, Bambi:2005fi} at $68\%$ c.l.\ As
we will shortly see, the large uncertainty on this bound makes it irrelevant for our analysis, therefore we do not implement this constraint in our likelihoods.\footnote{Very recently~\cite{Alvey:2019ctk} has updated this constraint, claiming that $\Delta G_N = 0.02 \pm 0.06$ at $95 \%$ C.L. While we postpone the application of this analysis to future work, we notice that such a constraint would affect our parameter space. However, as can be appreciated in Fig.\ \ref{triangleplot2}, the preferred value of $H_0$ should not be significantly reduced.} However, we take into account the effect of a varying Newton constant on the Helium abundance, by parametrizing it as an appropriate contribution to $\Delta N_{\text{eff}}$~, using the correspondence $\Delta G_N/G_N= -(7/43)~\Delta N_{\text{eff}}$ \cite{Iocco:2008va}.
We have performed two analyses of our varying $G_N$ model, shortly the $\Delta G_N$ model, comparing with the standard six-parameter $\Lambda$CDM model and the $\Delta N_{\rm eff}$ model. First, we assessed the Hubble tension in the $\Delta G_N$ model using early-time data only, that is Planck+BAO, without the inclusion of PN constraints. The fit to Planck+BAO gives $H_0=68.24_{-0.79}^{+0.5}$ km/s/Mpc,\footnote{The Gelman-Rubin~\cite{Gelman:1992zz} parameter $R -1$ reached less than 0.02, thus we considered our Monte Carlo chains to be well converged, according to the criterion~\cite{Gelman:1992zz} $R -1<0.1$.} so that the tension with SH$_{0}$ES 2019 is only marginally reduced, to the level of $3.8\sigma$.

\begin{figure}[t]
\vspace{-2cm}
	\includegraphics[width=\textwidth]{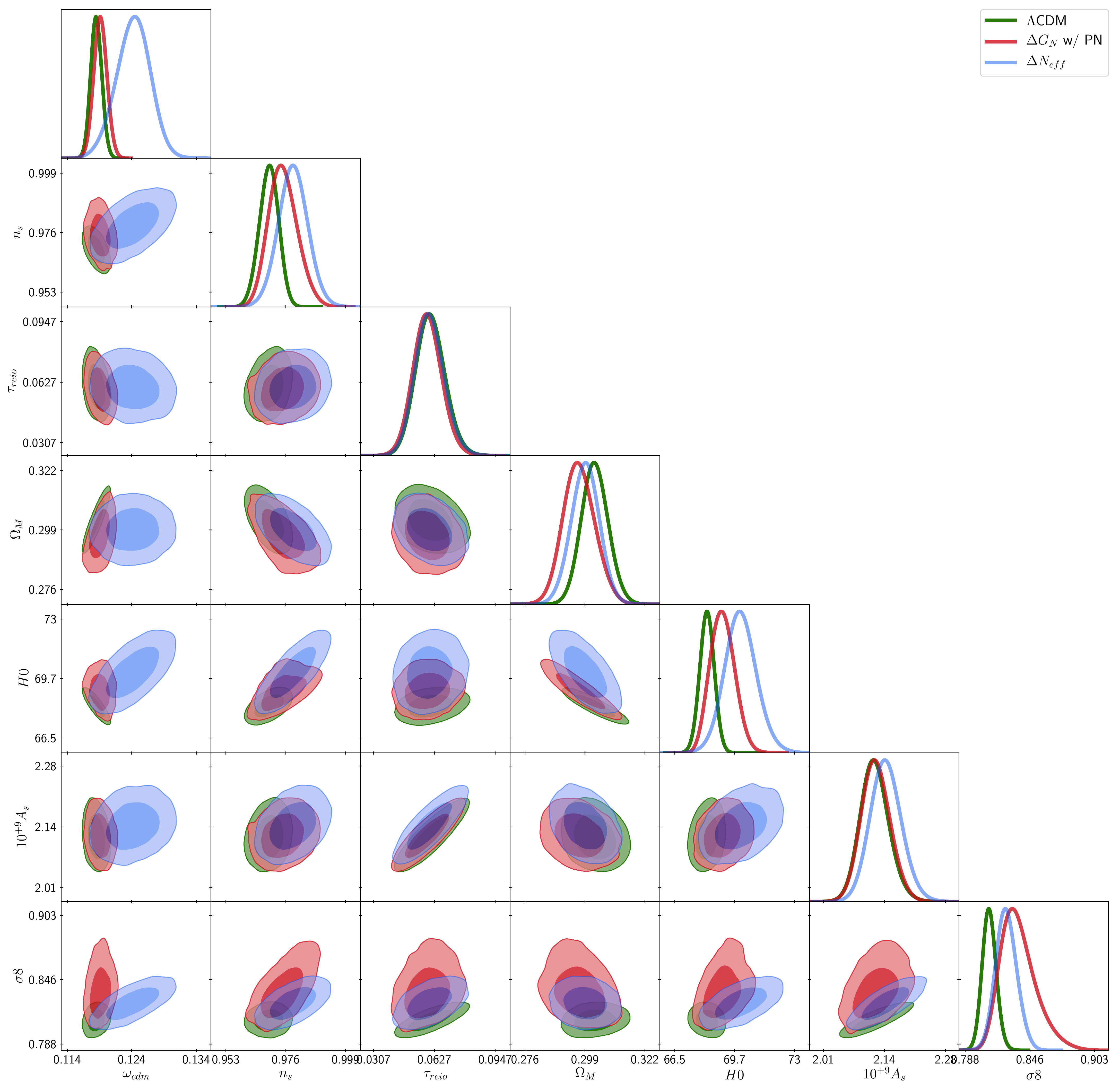}
	\caption{\small Constraints on parameters for our $\Delta G_N$ model vs.~the $\Delta N_{\rm eff}$ model and the base $\Lambda$CDM model, using
Planck 2018 high$-\ell$ TT,TE,EE+low$-\ell$ EE+ low$-\ell$ TT+lensing, BAO, Pantheon and SH$_{0}$ES 2019 data. PN constraints are included.
Parameters are our sampled MCMC parameters with flat priors. In particular a prior range has been set for the extra parameters: $-0.95<\beta \phi_I^2<0$
and $0<\phi_I<0.95$. Here $H_0$ is in km/s/Mpc. 
Contours contain $68 \%$ and $95 \% $ of the probability.  \label{triangleplot1}}
\end{figure}

\begin{figure}[t]
\centering
	\includegraphics[width=0.7\textwidth]{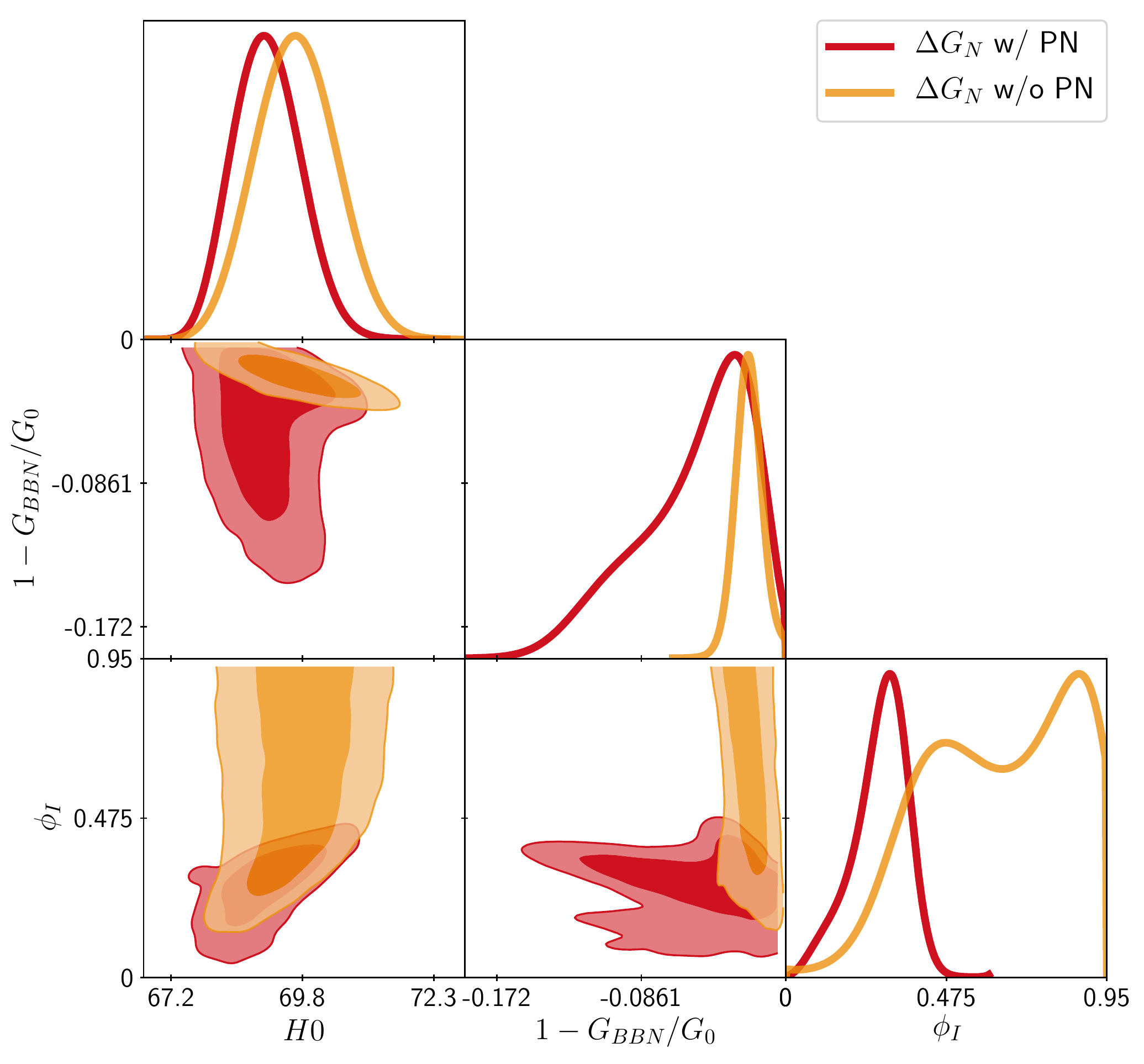}
	\caption{\small Above: constraints on parameters for our $\Delta G_N$ model, using
Planck 2018 high$-\ell$ TT,TE,EE+low$-\ell$ EE+ low$-\ell$ TT+lensing, BAO, Pantheon and SH$_{0}$ES 2019, with and without PN constraints. 
Parameters are our sampled MCMC parameters with flat priors. In particular a prior range has been set for the extra parameters: $-0.95<\beta \phi_I^2<0$
and $0<\phi_I<0.95$. Here $H_0$ is in km/s/Mpc  and $\phi_I$ is in Planck units.
Contours contain $68 \%$ and $95 \% $ of the probability.  \label{triangleplot3}}
\end{figure}

\begin{figure}[t]
\centering
\vspace{-2cm}
	\includegraphics[width=0.6\textwidth]{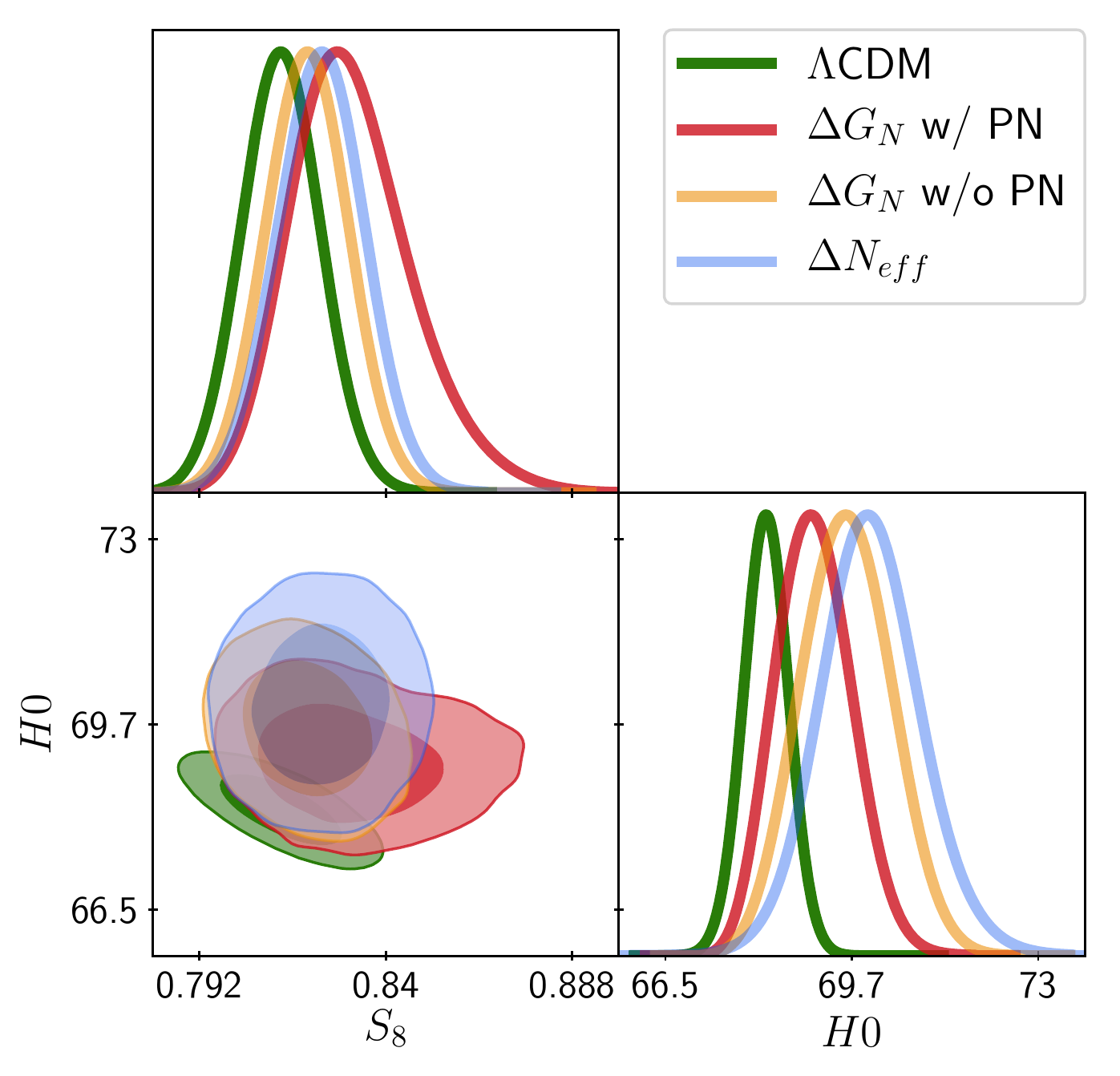}
	\caption{\small Constraints on the $S_8$ and $H_0$ parameters for the models considered in this work, using
Planck 2018 high$-\ell$ TT,TE,EE+low$-\ell$ EE+ low$-\ell$ TT+lensing, BAO, Pantheon and SH$_{0}$ES 2019 data, with and without PN constraints. 
Parameters are our sampled MCMC parameters with flat priors. Here $H_0$ is in km/s/Mpc. 
Contours contain $68 \%$ and $95 \% $ of the probability. For comparison, the recent combined analysis of~\cite{Heymans:2020gsg} finds $S_8=0.766^{+0.020}_{-0.014}$). \label{triangleplot4}}
\end{figure}

Interpreting the tension as a statistical fluctuation we combined the dataset above with Pantheon and SH$_{0}$ES 2019. In Table~\ref{tab:cosmoparametersl}  we show the comparison among the three models, with two options for the $\Delta G_N$ model (i.e. with and without PN constraints), and the contributions to the total $\chi^2$ are shown in Table~\ref{chi2}, for their best-fit values. Posterior distributions are shown in Figs.~\ref{triangleplot1}-\ref{triangleplot4}. 
The $\Delta N_{\rm eff}$ model gives the following results  at $68\%$ c.l.\ for Planck+BAO+PN+ SH$_{0}$ES:
\begin{eqnarray}
H_0&=&70.08_{-0.95}^{+0.91}  \text{km/s/Mpc} ,\nonumber \\
\Delta N_{\rm eff} &=& 0.34_{-0.16}^{+0.15} ,\nonumber \\
\Delta\chi^2&=&-3.4 \,
\end{eqnarray}
These results disfavor the pure $\Lambda$CDM value ($\Delta N_{\rm eff}=0$) at more than 2$\sigma$, which was not the case in the fit of~\cite{Aghanim:2018eyx}. This difference is due to the use of SH$_{0}$ES 2019 data, instead of  SH$_{0}$ES 2018. Using the Akaike Information Criterion~\cite{Akaike,Liddle:2007fy} one has $\Delta {\rm AIC}\equiv \Delta\chi^2-2\Delta p=-1.4$ in favor of the $\dn$ model,  where $\Delta p$ (equal to 1 in this case) is the number of additional parameters, beyond the six-parameter $\Lambda$CDM. For these runs the Gelman-Rubin parameter $R -1$ has reached less than $10^{-3}$.

Our scalar-tensor model including constraints on the PN parameter similarly improves by $\Delta\chi^2=-3.2$ compared to $\Lambda$CDM. However, in this case we find $H_0=69.08 _{-0.71}^{+0.6}$ km/s/Mpc, which is significantly smaller than in the $\Delta N_{\rm eff}$ model. Furthermore, this scenario is penalized by $\Delta {\rm AIC}\equiv \Delta\chi^2-2\Delta p=+0.82$, because it has two rather than one extra parameters. A small contribution to the  $\chi^2$ is  added to $\Lambda$CDM when comparing with this dataset, because of the PN constraint, which amounts to an additional $0.83$. For these runs the Gelman-Rubin parameter $R -1$ has reached less than 0.02. 
On the other hand, when removing PN constraint we find a more significant improvement of $\Delta\chi^2=-5.4$ compared to $\Lambda$CDM, with $\Delta {\rm AIC}\equiv \Delta\chi^2-2\Delta p=-1.4$. Furthermore, in this case we find $H_0=69.65^{+0.8}_{-0.78}$  km/s/Mpc, thus the model performs in this case very similarly to the $\Delta N_{\rm eff}$ model on the Hubble tension.

In Table~\ref{tab:cosmoparametersl} and in Fig.~\ref{triangleplot4} we also show the values of $S_8\equiv \sigma_8 \sqrt{\Omega_M/0.3}$, which is commonly used to compare with weak gravitational lensing of galaxies~\cite{Troxel:2017xyo} and galaxy clustering data~\cite{Abbott:2017wau}. These datasets give smaller values of $S_8$ than what is inferred by the CMB assuming $\Lambda$CDM (for instance, most recently~\cite{Heymans:2020gsg} finds $S_8=0.766^{+0.020}_{-0.014}$), and in general this discrepancy is exacerbated in models which address the $H_0$ tension. 
Very interestingly, we notice that in our model the value of $S_8$ is only very mildly shifted from its $\Lambda$CDM value when PN constraints are removed from the likelihood. The shift when PN constraints are included is more dramatic and is actually slightly larger than for the $\Delta N_{\text{eff}}$ model. Therefore, scenarios where PN constraints can be evaded might be particularly interesting for the $H_0$ tension, in that their tension with LSS datasets may turn out to be milder. 
When comparing with the EDE scenario of \cite{Poulin:2018cxd} with $n=3$ (recently reanalyzed from this perspective in \cite{Hill:2020osr}) we find smaller values of $S_8$ both with and without PN constraints, driven by a smaller value of $\Omega_{\text{c}} h^2\equiv \omega_c $ in our model than in EDE. Our $\Delta G_N$ model does not need an increase in $\omega_c$ with respect to $\Lambda$CDM, in contrast with the $\dn$ model, which needs more dark matter to keep the epoch of equality fixed, as can be seen in Fig.~\ref{triangleplot1} and \ref{triangleplot2}.

Let us now comment on previous related work. A model similar to the one considered in this paper was analyzed in~\cite{Rossi:2019lgt}, with the following differences: (1)  a cosmological constant  term was not included in $L_{\rm tot}$, whereas a potential $V(\phi)\propto f(\phi)^2$ was assumed instead;\footnote{In the Einstein frame such a potential would actually be rescaled by $f^{-2}(\phi)$ and thus act as a cosmological constant. 
}
(2) a fit of cosmological data including  SH$_{0}$ES was performed only for 
$\beta=-1/6$ (the conformal case); (3) the Post-Newtonian constraints were not included in the likelihood. Besides, the analysis in~\cite{Rossi:2019lgt} included older datasets (Planck 2015 instead of 2018, BOSS DR11 at $z_{\text{eff}}=0.57$ and $z_{\text{eff}}=0.32$ instead of BOSS DR12 and SH$_{0}$ES~2018 instead of SH$_{0}$ES~2019).

A Galileon model with varying $G_N$ for the Hubble tension has also been recently considered in \cite{Zumalacarregui:2020cjh}. It differs from ours in that the scalar field is exponentially coupled to the Ricci scalar (thus the coupling is effectively linear for small coupling rather than quadratic) and the Lagrangian features an extra term proportional to $X\, \square \phi$. Therefore, the model of \cite{Zumalacarregui:2020cjh} has one more extra parameter with respect to ours. Despite achieving a lower $\chi^2$ than our setup, the model appears to be in some tension with late-time constraints on modifications of General Relativity \cite{Zumalacarregui:2020cjh}.

\section{Conclusions} \label{Conclusions}

We have studied a very simple modification of gravity, where Newton's constant $G_N$ depends on a non-minimally coupled scalar field that decreases in time during the cosmological evolution, from an initial value $\phi_I$  to a final value $\phi_0\ll\phi_I$.
In such a varying $G_N$ model, the tension between Planck2018 + BAO data and the SH$_{0}$ES 2019 data is marginally reduced to at most $3.8\sigma$ (instead of $4.4\sigma$ in the case of Planck2018 assuming $\Lambda$CDM). This marginal improvement is similar to other models in the literature, which obtain larger values of $H_0$ only when supernovae data are included in the fit.

Our framework can be generically characterized by deviations from General Relativity at the Solar System scale. Therefore, the inclusion of existing constraints on Post-Newtonian parameters is important. Nevertheless, mechanisms and scenarios are known, in which such constraints can be evaded, by having different local and cosmological scalar field values. Keeping an open mind towards these latter scenarios, we have investigated both possibilities in this work.
Therefore, interpreting the above tension as a rare statistical fluctuation, we have performed a combined fit of  
Planck+BAO+Pantheon+SH$_{0}$ES data, with and without the inclusion of PN constraints. Comparing the best-fits with respect to the base $\Lambda$CDM, we have $\Delta\chi^2=-3.4$ for the $\dn$ model, with 1 extra parameter, which corresponds to a $\Delta {\rm AIC} =-1.4$ with the Akaike Information Criterion. The  $\Lambda$CDM limit of the $\dn$ model is disfavored at slightly more than 2$\sigma$, since we find $\Delta N_{\rm eff}=0.34_{-0.16}^{+0.15}$ and $H_0=70.08^{+0.91}_{-0.95}$.

For the $\Delta G_N$ model we find that local constraints on PN parameters importantly affect our conclusions. On the one hand, we find $\Delta\chi^2\simeq -3.2$ with respect to $\Lambda$CDM when PN constraints are included, and thus this scenario is penalized by $\Delta {\rm AIC}\simeq +0.8$. In this case we also find $H_0=69.08^{+0.6}_{-0.71}$. 
On the other hand, when PN constraints are removed, we find $\Delta\chi^2\simeq -5.4$ with respect to $\Lambda$CDM and thus $\Delta {\rm AIC}\simeq -1.4$. Furthermore, we also find larger values of $H_0=69.65^{+0.8}_{-0.78}$. From this point of view, scenarios where PN constraints could be neglected perform very similarly to the $\dn$ model.

Interestingly, we find that in our model the value of the $S_8$ parameter is only very mildly shifted with respect to its value in the $\Lambda$CDM model, when PN constraints are not included. This suggests that in this scenario the $S_8$ tension can be less relevant than in other popular frameworks for the $H_0$ tension, such as Early Dark Energy (EDE). We leave a more detailed investigation of this point for future work.

For scenarios where PN constraints are included, we find that the local field value is typically of order $10^{-4}-10^{-3}~M_p$ in the range of parameters preferred by the data. Therefore, a deviation from zero of the PN parameter $\gamma_{PN}-1$ is expected, constituting thus a potentially interesting prediction for proposed future experiments~\cite{Sakstein:2017pqi}, such as Phobos Laser Ranging~\cite{Turyshev:2010gk} that could go down to $10^{-7}$-$10^{-8}$ levels in $\gamma_{PN}-1$ and even to $10^{-9}$ with the LATOR~\cite{Plowman:2005fb,Turyshev:2007pt} and BEACON experiments~\cite{Turyshev:2008rh}, or to $2 \times 10^{-8}$ with gravitational time delay measurements (GTDM)~\cite{Ashby:2008lea}.
Present constraints on the coupling constant $\beta$ itself are irrelevant, since we only know that $\beta_0\approx-2\beta > -4.5$ from Pulsars~\cite{EspositoFarese:2004cc,Uzan:2010pm}.

Overall, we think that two aspects of our setup may be considered as an improvement over EDE solutions to the Hubble tension: first, our setup does not feature a coincidence problem on the onset of the scalar field dynamics; and second, it does not rely on non-generic scalar field potentials with several parameters, in contrast to \cite{Poulin:2018cxd, Smith:2019ihp}. 
We thus believe that these advantages may serve as a starting point for more sophisticated implementations of a varying $G_N$ to alleviate the Hubble tension.

\begin{table}[h!]
\vspace{0.5cm}
\centering
{\renewcommand{\arraystretch}{1.25} 
\resizebox{\textwidth}{!}{\begin{tabular}{|c | c | c | c | c |}
\hline
Parameter & $\Lambda$CDM & $\text{$\Delta G_N ~(\text{w/ PN})$}$ & $\text{$\Delta G_N~(\text{w/o PN})$}$ & $\Delta N_{\text{eff}}$ \\
\hline
$100~\omega_{b }$ & 2.254 (2.26) \small${}_{-0.014}^{+0.013}$ & 2.257  (2.26) \small${}_{-0.015}^{+0.015}$ & 2.247 (2.236) \small${}_{-0.014}^{+0.014}$  & 2.272 (2.262) \small${}^{+0.016}_{-0.016}$ \\
\hline
$\omega_{c }$ & 0.1183 (0.1189)  \small${}_{-0.00092}^{+0.00087}$ &   0.1189 (0.1188) \small${}_{-0.0011}^{+0.0011}$ & 0.1197 (0.1199) \small${}_{-0.0011}^{+0.0011}$ & 0.124 (0.1231) \small${}_{-0.0028}^{+0.0027}$ \\
\hline
$\tau_{\rm reio}$ & 0.06053 (0.06027) \small${}_{-0.0084}^{+0.0071}$ & 0.05895 (0.06037)  \small${}_{-0.008}^{+0.0073}$ & 0.05758 (0.05696) \small${}_{-0.008}^{+0.0069}$ & 0.06018 (0.05912) \small${}_{-0.0083}^{+0.0072}$ \\
\hline
$10^{+9} A_{\rm s} $ & 2.122 (2.123)   \small${}_{-0.035}^{+0.03}$ & 2.124 (2.126) \small${}_{-0.034}^{+0.031}$ & 2.12 (2.123)   \small${}_{-0.033}^{+0.029}$ & 2.147 (2.131) \small${}_{-0.036}^{+0.033}$\\
\hline
$n_{\rm s}$ & 0.9699 (0.9699)  \small${}_{-0.0036}^{+0.0038}$ & 0.9755 (0.9763) \small${}_{-0.0059}^{+0.0049} $ & 0.9708 (0.9714) \small${}_{-0.0042}^{+0.0038}$ & 0.9792 (0.9779) \small${}_{-0.0059}^{+0.0058}$ \\
\hline
$1-\frac{G^{\text{BBN}}_N}{G^0_N}$ or $\Delta N_{\text{eff}}$ & - & -0.05403 (-0.03226)\small${}_{-0.019}^{+0.044}$ & -0.02239 (-0.02339) \small${}_{-0.0087}^{+0.0082}$ & 0.3401 (0.2825) \small${}_{-0.16}^{+0.15}$\\
\hline
$\phi_I$ & - & 0.2789 (0.2662) \small${}_{-0.054}^{+0.097}$ & 0.6214 (0.6763) \small${}_{-0.11}^{+0.33}$ & - \\
\hline
$\sigma_8$ & 0.8097 (0.8119) \small${}_{-0.0066}^{+0.0061}$ &  0.8361 (0.8288) \small${}_{-0.018}^{+0.012}$  & 0.8292 (0.8327) \small${}_{-0.011}^{+0.011}$ & 0.825(0.821) \small${}_{-0.0095}^{+0.0095} $ \\
\hline
$\Omega_M$ & 0.3025 (0.3056) \small${}_{-0.0055}^{+0.0051}$ &  0.2966 (0.2972) \small${}_{-0.0064}^{+0.0062}$  &0.2932 (0.2919) \small${}_{-0.0068}^{+0.0065}$ & 0.2987 (0.3005) \small${}_{-0.0056}^{+0.0054}$ \\
\hline
$S_8$ & 0.8131 (0.8194) \small${}_{-0.01}^{+0.01} $ &  0.8312 (0.825) \small${}_{-0.018}^{+0.013}$  & 0.8198 (0.8213) \small${}_{-0.011}^{+0.011}$ & 0.8233 (0.8218) \small${}_{-0.011}^{+0.011}$ \\
\hline
$H_0$ [km/s/Mpc] & 68.23 (68.06)\small${}_{-0.41}^{+0.41}$ & 69.08 (68.97) \small${}_{-0.71}^{+0.6}$ & 69.65\ (69.83)  \small${}^{+0.8}_{-0.78}$ & 70.08 (69.64)  \small${}_{-0.95}^{+0.91}$ \\
\hline
\end{tabular}}
}
\caption{\small Mean values and $68\%$ confidence intervals for relevant cosmological parameters, obtained with the dataset ``Planck 2018 + BAO + Pantheon + S$H_0$ES 2019''. In parentheses, the best-fit values for each model. Cosmological parameters follow the standard notation  throughout the paper, as in~\cite{Aghanim:2018eyx}.}
\label{tab:cosmoparametersl}
\end{table}
\vspace{1cm}

\begin{table}[h!]
\centering
{\renewcommand{\arraystretch}{1.25} 
{\begin{tabular}{|c | c | c | c | c |}
\hline
Dataset & $\Lambda$CDM & $\text{$\Delta G_N~(\text{w/ PN})$}$ & $\text{$\Delta G_N~(\text{w/o PN})$}$ & $\Delta N_{\text{eff}}$ \\
\hline
 Planck highl TTTEEE  & 2352.18& 2354.55 & 2354.97 & 2357.80  \\
\hline
  Planck lowl EE   & 397.44 & 397.28 & 396.55 & 397.04\\
\hline
  Planck lowl TT  & 22.71 & 21.87 & 22.72 & 21.78\\
\hline
 Planck lensing  & 8.84 & 9.13 & 9.06 & 9.41\\
\hline
 Pantheon & 1027.06 & 1026.87 & 1026.92 & 1026.92\\
\hline
  SH$_0$ES 2019 & 17.71 & 13.34 & 7.23 & 9.56\\
\hline
 bao boss dr12 & 3.81 & 3.45 & 4.08 &  3.46\\ 
\hline
  bao smallz 2014 & 1.45 & 2.16 & 2.74 & 1.80\\
\hline
  PN  & (0.83) & 0.83 &  (0.83)  & (0.83) \\ 
\hline
Total  & 3831.19 & 3828.84 & 3825.79 &  3827.78\\
\hline
\hline
$\Delta \chi^2$ & 0 & $-3.18$ & $-5.4$ &  $-3.41$\\
\hline
$\Delta$AIC & 0 & +0.82 & $-1.4$ &  $-1.4$\\
\hline
\end{tabular}}
}
\caption{\small Contributions to the total $\chi^2_{\rm eff}$ for individual datasets, for the best-fits of $\Lambda$CDM, $\Delta G_N$ and $\Delta N_{\rm eff}$ models. The $\Delta \chi^2$ and $\Delta$ AIC for the $G_N$ model with PN constraints with respect to $\Lambda$CDM has been obtained by adding the PN contribution to the $\Lambda$CDM model as well.   \label{chi2}}
\end{table}
	


\emph{\textbf{Acknowledgments:}} {\small 
	We thank  Emilio Bellini, Nicola Bellomo, Alberto Belloni, Samuel Brieden, Wilmar Cardona, Ivan Esteban, Julien Lesgourgues,  Francesco Montanari, Savvas Nesseris, Jordi Salvad\'o, Ignacy Sawicki, Pasquale Serpico, Joan Sol\`a Peracaula, Caterina Umilt\`a and Licia Verde for useful discussions and help with codes and Ricardo Z. Ferreira for collaboration in the early stages of this work. F.R. thanks Marianna Annunziatella for help with installations of numerical codes.  We acknowledge the use of Hydra Cluster at IFT, Madrid and of Tufts HPC Research Cluster. A.N. is grateful  to the Physics Department of the University of Padova for the hospitality, while this work was completed. This work is supported by the grants FPA2016-76005-C2-2-P, MDM-2014-0369 of ICCUB (Unidad
de Excelencia Maria de Maeztu), AGAUR2017-SGR-754. The work of G.B. is funded by a {\it Contrato de Atracci\'on de Talento (Modalidad 1) de la Comunidad de Madrid} (Spain), with number 2017-T1/TIC-5520, by {\it MINECO} (Spain) under contract FPA2016-78022-P, {\it MCIU} (Spain) through contract PGC2018-096646-A-I00 and by the IFT UAM-CSIC Centro de Excelencia Severo Ochoa SEV-2016-0597 grant. G.B. thanks the Leinweber Center for Theoretical Physics of the University of Michigan and KITP for hospitality  and acknowledges support from the NSF, under Grant No. NSF-1748958.}	
	
%

\newpage

\providecommand{\href}[2]{#2}\begingroup\raggedright\endgroup


\begin{thebibliography}{10}

\bibitem{Riess:2019cxk}
A.~G. Riess, S.~Casertano, W.~Yuan, L.~M. Macri and D.~Scolnic, \emph{{Large
  Magellanic Cloud Cepheid Standards Provide a 1\% Foundation for the
  Determination of the Hubble Constant and Stronger Evidence for Physics beyond
  $\Lambda$CDM}},
  \href{https://doi.org/10.3847/1538-4357/ab1422}{\emph{Astrophys. J.}
  {\bfseries 876} (2019) 85}
  [\href{https://arxiv.org/abs/1903.07603}{{\ttfamily 1903.07603}}].

\bibitem{Aghanim:2018eyx}
{\scshape Planck} collaboration, N.~Aghanim et~al., \emph{{Planck 2018 results.
  VI. Cosmological parameters}},
  \href{https://arxiv.org/abs/1807.06209}{{\ttfamily 1807.06209}}.

  \bibitem{Birrer:2018vtm}
S.~Birrer et~al., \emph{{H0LiCOW - IX. Cosmographic analysis of the doubly
  imaged quasar SDSS 1206+4332 and a new measurement of the Hubble constant}},
  \href{https://doi.org/10.1093/mnras/stz200}{\emph{Mon. Not. Roy. Astron.
  Soc.} {\bfseries 484} (2019) 4726}
  [\href{https://arxiv.org/abs/1809.01274}{{\ttfamily 1809.01274}}].
  
  \bibitem{Freedman:2019jwv}
W.~L. Freedman et~al., \emph{{The Carnegie-Chicago Hubble Program. VIII. An
  Independent Determination of the Hubble Constant Based on the Tip of the Red
  Giant Branch}},  [\href{https://arxiv.org/abs/1907.05922}{{\ttfamily
  1907.05922}}].
  

\bibitem{Verde:2019ivm}
L.~Verde, T.~Treu and A.~G. Riess, \emph{{Tensions between the Early and the
  Late Universe}},   \href{https://doi.org/10.1038/s41550-019-0902-0}{\emph{{Nature Astronomy 2019}}, 2019},
  [\href{https://arxiv.org/abs/1907.10625}{{\ttfamily 1907.10625}}].




  
\bibitem{Riess:2020sih}
A.~G.~Riess, \emph{{The Expansion of the Universe is Faster than Expected}}, \href{https://doi.org/10.1038/s42254-019-0137-0}{Nature Rev. Phys. \textbf{2} (2019) no.1, 10-12}, [\href{https://arxiv.org/abs/2001.03624}{\ttfamily 2001.03624}].


\bibitem{Knox:2019rjx}
L.~Knox and M.~Millea, \emph{{The Hubble Hunter's Guide}},
  \href{https://arxiv.org/abs/1908.03663}{{\ttfamily 1908.03663}}.

\bibitem{Aylor:2018drw}
K.~Aylor, M.~Joy, L.~Knox, M.~Millea, S.~Raghunathan and W.~L.~K. Wu,
  \emph{{Sounds Discordant: Classical Distance Ladder \& $\Lambda$CDM -based
  Determinations of the Cosmological Sound Horizon}},
  \href{https://doi.org/10.3847/1538-4357/ab0898}{\emph{Astrophys. J.}
  {\bfseries 874} (2019) 4} [\href{https://arxiv.org/abs/1811.00537}{{\ttfamily
  1811.00537}}].
  


\bibitem{DEramo:2018vss}
F.~D'Eramo, R.~Z. Ferreira, A.~Notari and J.~L. Bernal, \emph{{Hot Axions and
  the $H_0$ tension}},
  \href{https://doi.org/10.1088/1475-7516/2018/11/014}{\emph{JCAP} {\bfseries
  1811} (2018) 014} [\href{https://arxiv.org/abs/1808.07430}{{\ttfamily
  1808.07430}}].
  
  \bibitem{Alcaniz:2019kah}
J.~Alcaniz, N.~Bernal, A.~Masiero and F.~S.~Queiroz,
\emph{{Light Dark Matter: A Common Solution to the Lithium and ${H_0}$ Problems}}
[\href{https://arxiv.org/abs/1912.05563}{{\ttfamily 1912.05563}}].
  
  \bibitem{Graef:2018fzu}
L.~L.~Graef, M.~Benetti and J.~S.~Alcaniz,
\emph{{Primordial gravitational waves and the H0-tension problem}}
\href{https://journals.aps.org/prd/abstract/10.1103/PhysRevD.99.043519}{\emph{Phys. Rev. D} \textbf{99} (2019) no.4, 043519}
  [\href{https://arxiv.org/abs/1809.04501}{{\ttfamily 1809.04501}}].

\bibitem{Poulin:2018cxd}
V.~Poulin, T.~L. Smith, T.~Karwal and M.~Kamionkowski, \emph{{Early Dark Energy
  Can Resolve The Hubble Tension}},
  \href{https://doi.org/10.1103/PhysRevLett.122.221301}{\emph{Phys. Rev. Lett.}
  {\bfseries 122} (2019) 221301}
  [\href{https://arxiv.org/abs/1811.04083}{{\ttfamily 1811.04083}}].

\bibitem{Smith:2019ihp}
T.~L. Smith, V.~Poulin and M.~A. Amin, \emph{{Oscillating scalar fields and the
  Hubble tension: a resolution with novel signatures}},
  \href{https://arxiv.org/abs/1908.06995}{{\ttfamily 1908.06995}}.

\bibitem{Lin:2019qug}
M.-X. Lin, G.~Benevento, W.~Hu and M.~Raveri, \emph{{Acoustic Dark Energy:
  Potential Conversion of the Hubble Tension}},
  \href{https://doi.org/10.1103/PhysRevD.100.063542}{\emph{Phys. Rev.}
  {\bfseries D100} (2019) 063542}
  [\href{https://arxiv.org/abs/1905.12618}{{\ttfamily 1905.12618}}].

\bibitem{Agrawal:2019lmo}
P.~Agrawal, F.-Y. Cyr-Racine, D.~Pinner and L.~Randall, \emph{{Rock 'n' Roll
  Solutions to the Hubble Tension}},
  \href{https://arxiv.org/abs/1904.01016}{{\ttfamily 1904.01016}}.

\bibitem{Hill:2020osr}
J.~C. Hill, E.~McDonough, M.~W. Toomey and S.~Alexander, \emph{{Early Dark
  Energy Does Not Restore Cosmological Concordance}},
  \href{https://arxiv.org/abs/2003.07355}{{\ttfamily 2003.07355}}.

\bibitem{Iocco:2008va}
F.~Iocco, G.~Mangano, G.~Miele, O.~Pisanti and P.~D. Serpico, \emph{{Primordial
  Nucleosynthesis: from precision cosmology to fundamental physics}},
  \href{https://doi.org/10.1016/j.physrep.2009.02.002}{\emph{Phys.\ Rept.}
  {\bfseries 472} (2009) 1} [\href{https://arxiv.org/abs/0809.0631}{{\ttfamily
  0809.0631}}].

\bibitem{Brans:1961sx}
C.~Brans and R.~H. Dicke, \emph{{Mach's principle and a relativistic theory of
  gravitation}}, \href{https://doi.org/10.1103/PhysRev.124.925}{\emph{Phys.
  Rev.} {\bfseries 124} (1961) 925}.
  
\bibitem{Vainshtein:1972sx}
A.~I.~Vainshtein,
Phys. Lett. B \textbf{39} (1972), 393-394
\href{doi:10.1016/0370-2693(72)90147-5}.

\bibitem{Khoury:2003aq}
J.~Khoury and A.~Weltman,
Phys. Rev. Lett. \textbf{93} (2004), 171104
doi:10.1103/PhysRevLett.93.171104
[\href{https://arXiv:astro-ph/0309300}{{\ttfamily
  astro-ph/0309300}}]].
  
\bibitem{Lin:2018nxe}
M.~Lin, M.~Raveri and W.~Hu,
\emph{{Phenomenology of Modified Gravity at Recombination}},
\href{doi:10.1103/PhysRevD.99.043514}{\emph{Phys. Rev.} {\bfseries{D99}} (2019) 4}
[\href{https://arxiv.org/abs/1810.02333}{{\ttfamily
  1810.02333}}].

\bibitem{Rossi:2019lgt}
M.~Rossi, M.~Ballardini, M.~Braglia, F.~Finelli, D.~Paoletti, A.~A. Starobinsky
  et~al., \emph{{Cosmological constraints on post-Newtonian parameters in
  effectively massless scalar-tensor theories of gravity}},
  \href{https://doi.org/10.1103/PhysRevD.100.103524}{\emph{Phys. Rev.}
  {\bfseries D100} (2019) 103524}
  [\href{https://arxiv.org/abs/1906.10218}{{\ttfamily 1906.10218}}].
  
\bibitem{Umilta:2015cta}
C.~Umilt\'a, M.~Ballardini, F.~Finelli and D.~Paoletti,
\emph{{CMB and BAO constraints for an induced gravity dark energy model with a quartic potential}},
\href{https://iopscience.iop.org/article/10.1088/1475-7516/2015/08/017}{\emph{
JCAP \textbf{08} (2015), 017}}
  [\href{https://arxiv.org/abs/1507.00718}{{\ttfamily 1507.00718}}].

\bibitem{Ballardini:2016cvy}
M.~Ballardini, F.~Finelli, C.~Umilt\'a and D.~Paoletti,
\emph{{Cosmological constraints on induced gravity dark energy models}},
\href{https://iopscience.iop.org/article/10.1088/1475-7516/2016/05/067/meta}{\emph{
JCAP \textbf{05} (2016), 067}}
[\href{https://arxiv.org/abs/1601.03387}{{\ttfamily 1601.03387}}].

  
\bibitem{Sola:2019jek}
J.~Sol\`a Peracaula, A.~Gomez-Valent, J.~de Cruz P\'erez and C.~Moreno-Pulido,
\emph{Brans Dicke Gravity with a Cosmological Constant Smoothes Out $\Lambda$CDM Tensions}
\href{doi:10.3847/2041-8213/ab53e9}{\emph{Astrophys. J.} {\bfseries 886} (2019) 1}
[\href{https://arxiv.org/abs/1909.02554}{{\ttfamily 1909.02554}}].

\bibitem{Zumalacarregui:2020cjh}
M.~Zumalacarregui, \emph{{Gravity in the Era of Equality: Towards solutions to
  the Hubble problem without fine-tuned initial conditions}},
  \href{https://arxiv.org/abs/2003.06396}{{\ttfamily 2003.06396}}.

\bibitem{Sakstein:2019fmf}
J.~Sakstein and M.~Trodden, \emph{{Early dark energy from massive neutrinos --
  a natural resolution of the Hubble tension}},
  \href{https://arxiv.org/abs/1911.11760}{{\ttfamily 1911.11760}}.
  
  
\bibitem{Hart:2019dxi}
L.~Hart and J.~Chluba,
\emph{Updated fundamental constant constraints from Planck 2018 data and possible relations to the Hubble tension},
\href{doi:10.1093/mnras/staa412}{\emph{Mon. Not. Roy. Astron. Soc.} {\bfseries 493} (2020) 3}
[\href{https://arxiv.org/abs/1912.03986}{{\ttfamily 1912.03986}}].

\bibitem{Damour:1994zq}
T.~Damour and A.~M. Polyakov, \emph{{The String dilaton and a least coupling
  principle}}, \href{https://doi.org/10.1016/0550-3213(94)90143-0}{\emph{Nucl.
  Phys.} {\bfseries B423} (1994) 532}
  [\href{https://arxiv.org/abs/hep-th/9401069}{{\ttfamily hep-th/9401069}}].

\bibitem{Cembranos:2009ds}
J.~A.~R. Cembranos, K.~A. Olive, M.~Peloso and J.-P. Uzan, \emph{{Quantum
  Corrections to the Cosmological Evolution of Conformally Coupled Fields}},
  \href{https://doi.org/10.1088/1475-7516/2009/07/025}{\emph{JCAP} {\bfseries
  0907} (2009) 025} [\href{https://arxiv.org/abs/0905.1989}{{\ttfamily
  0905.1989}}].

\bibitem{Zumalacarregui:2016pph}
M.~Zumalacarregui, E.~Bellini, I.~Sawicki, J.~Lesgourgues and P.~G. Ferreira,
  \emph{{hi class: Horndeski in the Cosmic Linear Anisotropy Solving System}},
  \href{https://doi.org/10.1088/1475-7516/2017/08/019}{\emph{JCAP} {\bfseries
  1708} (2017) 019} [\href{https://arxiv.org/abs/1605.06102}{{\ttfamily
  1605.06102}}].

\bibitem{Bellini:2019syt}
E.~Bellini, I.~Sawicki and M.~Zumalacarregui, \emph{{hi class: Background
  Evolution, Initial Conditions and Approximation Schemes}},
  \href{https://arxiv.org/abs/1909.01828}{{\ttfamily 1909.01828}}.

\bibitem{Blas_2011}
D.~Blas, J.~Lesgourgues and T.~Tram, \emph{The cosmic linear anisotropy solving
  system (class). part ii: Approximation schemes},
  \href{https://doi.org/10.1088/1475-7516/2011/07/034}{\emph{Journal of
  Cosmology and Astroparticle Physics} {\bfseries 2011} (2011) 034?034}.
  
\bibitem{Scolnic:2017caz}
D.~M.~Scolnic, D.~O.~Jones, A.~Rest, Y.~C.~Pan, R.~Chornock, R.~J.~Foley, M.~E.~Huber, R.~Kessler, G.~Narayan, A.~G.~Riess, S.~Rodney, E.~Berger, D.~J.~Brout, P.~J.~Challis, M.~Drout, D.~Finkbeiner, R.~Lunnan, R.~P.~Kirshner, N.~E.~Sanders, E.~Schlafly, S.~Smartt, C.~W.~Stubbs, J.~Tonry, W.~M.~Wood-Vasey, M.~Foley, J.~Hand, E.~Johnson, W.~S.~Burgett, K.~C.~Chambers, P.~W.~Draper, K.~W.~Hodapp, N.~Kaiser, R.~P.~Kudritzki, E.~A.~Magnier, N.~Metcalfe, F.~Bresolin, E.~Gall, R.~Kotak, M.~McCrum and K.~W.~Smith,
Astrophys. J. \textbf{859} (2018) no.2, 101
\href{doi:10.3847/1538-4357/aab9bb}
[\href{https://arXiv.org/abs/1710.00845}{{\ttfamily 1710.00845}} ].

\bibitem{Aghanim:2019ame}
{\scshape Planck} collaboration, N.~Aghanim et~al., \emph{{Planck 2018 results.
  V. CMB power spectra and likelihoods}},
  \href{https://arxiv.org/abs/1907.12875}{{\ttfamily 1907.12875}}.

\bibitem{Beutler:2011hx}
F.~Beutler, C.~Blake, M.~Colless, D.~H. Jones, L.~Staveley-Smith, L.~Campbell
  et~al., \emph{{The 6dF Galaxy Survey: Baryon Acoustic Oscillations and the
  Local Hubble Constant}},
  \href{https://doi.org/10.1111/j.1365-2966.2011.19250.x}{\emph{Mon. Not. Roy.
  Astron. Soc.} {\bfseries 416} (2011) 3017}
  [\href{https://arxiv.org/abs/1106.3366}{{\ttfamily 1106.3366}}].

\bibitem{Ross:2014qpa}
A.~J. Ross, L.~Samushia, C.~Howlett, W.~J. Percival, A.~Burden and M.~Manera,
  \emph{{The clustering of the SDSS DR7 main Galaxy sample ? I. A 4 per cent
  distance measure at $z = 0.15$}},
  \href{https://doi.org/10.1093/mnras/stv154}{\emph{Mon. Not. Roy. Astron.
  Soc.} {\bfseries 449} (2015) 835}
  [\href{https://arxiv.org/abs/1409.3242}{{\ttfamily 1409.3242}}].

\bibitem{Alam:2016hwk}
{\scshape BOSS} collaboration, S.~Alam et~al., \emph{{The clustering of
  galaxies in the completed SDSS-III Baryon Oscillation Spectroscopic Survey:
  cosmological analysis of the DR12 galaxy sample}},
  \href{https://doi.org/10.1093/mnras/stx721}{\emph{Mon. Not. Roy. Astron.
  Soc.} {\bfseries 470} (2017) 2617}
  [\href{https://arxiv.org/abs/1607.03155}{{\ttfamily 1607.03155}}].

\bibitem{Audren:2012wb}
B.~Audren, J.~Lesgourgues, K.~Benabed and S.~Prunet, \emph{{Conservative
  Constraints on Early Cosmology: an illustration of the Monte Python
  cosmological parameter inference code}},
  \href{https://doi.org/10.1088/1475-7516/2013/02/001}{\emph{JCAP} {\bfseries
  1302} (2013) 001} [\href{https://arxiv.org/abs/1210.7183}{{\ttfamily
  1210.7183}}].

\bibitem{EspositoFarese:2004cc}
G.~Esposito-Farese, \emph{{Tests of scalar-tensor gravity}},
  \href{https://doi.org/10.1063/1.1835173}{\emph{AIP Conf. Proc.} {\bfseries
  736} (2004) 35} [\href{https://arxiv.org/abs/gr-qc/0409081}{{\ttfamily
  gr-qc/0409081}}].

\bibitem{Bertotti:2003rm}
B.~Bertotti, L.~Iess and P.~Tortora, \emph{{A test of general relativity using
  radio links with the Cassini spacecraft}},
  \href{https://doi.org/10.1038/nature01997}{\emph{Nature} {\bfseries 425}
  (2003) 374}.

\bibitem{Fienga:2014bvy}
A.~Fienga, J.~Laskar, P.~Exertier, H.~Manche and M.~Gastineau, \emph{{Tests of
  General relativity with planetary orbits and Monte Carlo simulations}},
  \href{https://arxiv.org/abs/1409.4932}{{\ttfamily 1409.4932}}.

\bibitem{Boisseau:2000pr}
B.~Boisseau, G.~Esposito-Farese, D.~Polarski and A.~A. Starobinsky,
  \emph{{Reconstruction of a scalar tensor theory of gravity in an accelerating
  universe}}, \href{https://doi.org/10.1103/PhysRevLett.85.2236}{\emph{Phys.
  Rev. Lett.} {\bfseries 85} (2000) 2236}
  [\href{https://arxiv.org/abs/gr-qc/0001066}{{\ttfamily gr-qc/0001066}}].

\bibitem{Copi:2003xd}
C.~J. Copi, A.~N. Davis and L.~M. Krauss, \emph{{A New nucleosynthesis
  constraint on the variation of G}},
  \href{https://doi.org/10.1103/PhysRevLett.92.171301}{\emph{Phys. Rev. Lett.}
  {\bfseries 92} (2004) 171301}
  [\href{https://arxiv.org/abs/astro-ph/0311334}{{\ttfamily
  astro-ph/0311334}}].

\bibitem{Bambi:2005fi}
C.~Bambi, M.~Giannotti and F.~L. Villante, \emph{{The Response of primordial
  abundances to a general modification of G(N) and/or of the early Universe
  expansion rate}},
  \href{https://doi.org/10.1103/PhysRevD.71.123524}{\emph{Phys. Rev.}
  {\bfseries D71} (2005) 123524}
  [\href{https://arxiv.org/abs/astro-ph/0503502}{{\ttfamily
  astro-ph/0503502}}].

\bibitem{Alvey:2019ctk}
J.~Alvey, N.~Sabti, M.~Escudero and M.~Fairbairn, \emph{{Improved BBN
  Constraints on the Variation of the Gravitational Constant}},
  \href{https://arxiv.org/abs/1910.10730}{{\ttfamily 1910.10730}}.

\bibitem{Gelman:1992zz}
A.~Gelman and D.~B. Rubin, \emph{{Inference from Iterative Simulation Using
  Multiple Sequences}},
  \href{https://doi.org/10.1214/ss/1177011136}{\emph{Statist. Sci.} {\bfseries
  7} (1992) 457}.

\bibitem{Akaike}
H.~Akaike, \emph{{Information criteria for astrophysical model selection}},
  {\emph{IEEE T. Automat. Contr.} {\bfseries 19, 716} (1974) }.

\bibitem{Liddle:2007fy}
A.~R. Liddle, \emph{{Information criteria for astrophysical model selection}},
  \href{https://doi.org/10.1111/j.1745-3933.2007.00306.x}{\emph{Mon. Not. Roy.
  Astron. Soc.} {\bfseries 377} (2007) L74}
  [\href{https://arxiv.org/abs/astro-ph/0701113}{{\ttfamily
  astro-ph/0701113}}].

\bibitem{Troxel:2017xyo}
{\scshape DES} collaboration, M.~A. Troxel et~al., \emph{{Dark Energy Survey
  Year 1 results: Cosmological constraints from cosmic shear}},
  \href{https://doi.org/10.1103/PhysRevD.98.043528}{\emph{Phys. Rev.}
  {\bfseries D98} (2018) 043528}
  [\href{https://arxiv.org/abs/1708.01538}{{\ttfamily 1708.01538}}].

\bibitem{Abbott:2017wau}
{\scshape DES} collaboration, T.~M.~C. Abbott et~al., \emph{{Dark Energy Survey
  year 1 results: Cosmological constraints from galaxy clustering and weak
  lensing}}, \href{https://doi.org/10.1103/PhysRevD.98.043526}{\emph{Phys.
  Rev.} {\bfseries D98} (2018) 043526}
  [\href{https://arxiv.org/abs/1708.01530}{{\ttfamily 1708.01530}}].
  
\bibitem{Heymans:2020gsg}
C.~Heymans, T.~Troester, M.~Asgari, C.~Blake, H.~Hildebrandt, B.~Joachimi, K.~Kuijken, C.~A.~Lin, A.~G.~Sanchez, J.~L.~v.~Busch, A.~H.~Wright, A.~Amon, M.~Bilicki, J.~de Jong, M.~Crocce, A.~Dvornik, T.~Erben, F.~Getman, B.~Giblin, K.~Glazebrook, H.~Hoekstra, S.~Joudaki, A.~Kannawadi, C.~Lidman, F.~Koehlinger, L.~Miller, N.~R.~Napolitano, D.~Parkinson, P.~Schneider, H.~Shan and C.~Wolf,
\emph{KiDS-1000 Cosmology: Multi-probe weak gravitational lensing and spectroscopic galaxy clustering constraints},
\href{https://arxiv.org/abs/2007.15632} {{\ttfamily 2007.15632}}.

\bibitem{Sakstein:2017pqi}
J.~Sakstein, \emph{{Tests of Gravity with Future Space-Based Experiments}},
  \href{https://doi.org/10.1103/PhysRevD.97.064028}{\emph{Phys. Rev.}
  {\bfseries D97} (2018) 064028}
  [\href{https://arxiv.org/abs/1710.03156}{{\ttfamily 1710.03156}}].

\bibitem{Turyshev:2010gk}
S.~G. Turyshev, W.~Farr, W.~M. Folkner, A.~R. Girerd, H.~Hemmati, T.~W. Murphy,
  Jr. et~al., \emph{{Advancing Tests of Relativistic Gravity via Laser Ranging
  to Phobos}}, \href{https://doi.org/10.1007/s10686-010-9199-9}{\emph{Exper.
  Astron.} {\bfseries 28} (2010) 209}
  [\href{https://arxiv.org/abs/1003.4961}{{\ttfamily 1003.4961}}].

\bibitem{Plowman:2005fb}
J.~E. Plowman and R.~W. Hellings, \emph{{LATOR covariance analysis}},
  \href{https://doi.org/10.1088/0264-9381/23/2/002}{\emph{Class. Quant. Grav.}
  {\bfseries 23} (2006) 309}
  [\href{https://arxiv.org/abs/gr-qc/0505064}{{\ttfamily gr-qc/0505064}}].

\bibitem{Turyshev:2007pt}
S.~G. Turyshev and M.~Shao, \emph{{The Laser Astrometric Test of Relativity:
  Science, Technology, and Mission Design}},
  \href{https://doi.org/10.1142/S0218271807011747}{\emph{Int. J. Mod. Phys.}
  {\bfseries D16} (2007) 2191}
  [\href{https://arxiv.org/abs/gr-qc/0701102}{{\ttfamily gr-qc/0701102}}].

\bibitem{Turyshev:2008rh}
S.~G. Turyshev, B.~Lane, M.~Shao and A.~Girerd, \emph{{A Search for New Physics
  with the BEACON Mission}},
  \href{https://doi.org/10.1142/S0218271809014893}{\emph{Int. J. Mod. Phys.}
  {\bfseries D18} (2009) 1025}
  [\href{https://arxiv.org/abs/0805.4033}{{\ttfamily 0805.4033}}].

\bibitem{Ashby:2008lea}
N.~Ashby and P.~L. Bender, \emph{{Measurement of the Shapiro Time Delay Between
  Drag-Free Spacecraft}},
  \href{https://doi.org/10.1007/978-3-540-34377-6_10}{\emph{Astrophys. Space
  Sci. Libr.} {\bfseries 349} (2008) 219}.

\bibitem{Uzan:2010pm}
J.-P. Uzan, \emph{{Varying Constants, Gravitation and Cosmology}},
  \href{https://doi.org/10.12942/lrr-2011-2}{\emph{Living Rev. Rel.} {\bfseries
  14} (2011) 2} [\href{https://arxiv.org/abs/1009.5514}{{\ttfamily
  1009.5514}}].

\end{thebibliography}
	
\end{document}